\documentclass{article}     

\usepackage{graphicx}
\usepackage{dcolumn}
\usepackage{bm}

\usepackage[utf8]{inputenc}
\usepackage[T1]{fontenc}
\usepackage{mathptmx}
\usepackage{amsfonts}
\usepackage{bbm}
\usepackage[bottom]{footmisc}
\usepackage[utf8]{inputenc}

\begin{document}

\title{Conceptual variables, quantum theory, and statistical inference theory}



\author{Inge S. Helland \\ Department of Mathematics, University of Oslo, \\ P.O.Box 1053 Blindern, N-0316 Oslo, Norway\\ingeh@math.uio.no}






\date{}

\maketitle


\begin{abstract}
A different approach towards quantum theory is proposed in this paper. The basis is taken to be conceptual variables, physical variables that may be accessible or inaccessible, i.e., it may be possible or impossible to assign numerical values to them. In an epistemic process, the accessible variables are just ideal observations as observed by an actor or by some communicating actors. Group actions are defined on these variables, and using group representation theory this is the basis for developing the Hilbert space formalism here. Operators corresponding to accessible conceptual variables are derived as a result of the formalism, and in the discrete case it is argued that the possible physical values are the eigenvalues of these operators. The Born formula is derived under specific assumptions. The whole discussion here is a supplement to the author's book [1]. The interpretation of quantum states (or eigenvector spaces) implied by this approach is as focused questions to nature together with sharp answers to those questions. Resolutions if the identity are then connected to the questions themselves; these may be complementary in the sense defined by Bohr. This interpretation may be called a general epistemic interpretation of quantum theory. It is similar to Zwirn's recent Convival Solipsism, and also to QBism, and more generally, can be seen as a concrete implementation of Rovelli's Relational Quantum Mechanics. The focus in the present paper is, however, as much on foundation as on interpretation. But the simple consequences of an epistemic interpretation for some so called quantum paradoxes are discussed. Connections to statistical inference theory are discussed in a preliminary way, both through an example and through a brief discussion of quantum measurement theory.


\end{abstract}

\section{Introduction}
\label{intro}

One purpose of this article is to pave the way towards discussing possible connections between quantum foundation and aspects of the foundation of statistical inference theory. The final goal is to find links between these two scientific areas, but also to point at differences. For this we need a common language, here expressed by conceptual variables, physical variables attached to some agent. Another purpose of the article is to give an introduction to the approach towards formal quantum theory through such variables. During these developments, I will discuss essential aspects both of the foundation of quantum theory and of the interpretation of the theory.

For the quantum theory discussion, take as a point of departure the famous Bell experiment, discussed by many authors; here I rely mainly on the discussion in Gill [2].

Alice and Bob have a space-like separation and thus  cannot communicate. They each measure the spin component of a particle from an entangled pair. The measurement $\theta$ made by Alice is what I call a conceptual variable, attached to her. Similarly, Bob has a measurement, his conceptual variable $\psi$. It is crucial for me that there do exist experiments where one has to attach a conceptual variable to an observer.

To define these conceptual variables, I will think of ideal experiments. In my opinion, for non-ideal, imperfect measurement apparata, one should use statistical modeling. For this reason and for several other reasons I include a section on the foundation of statistical inference in this paper.

Now each of Alice and Bob has a free will, and they choose some setting for their measurement. The hypothetical (perfect) measurement results $A$ or $A'$ for Alice, given her settings $a$ or $a'$, and for Bob, the hypothetical measurements $B$ and $B'$, given his settings $b$ or $b'$ are the background for one of Bell's inequalities, the CHSH inequality:
\begin{equation}
E(AB)+E(AB')+E(A'B)-E(A'B')\le 2,
\label{Bell}
\end{equation}
which can be found formally by taking expectations term by term in the inequality
\begin{equation}
AB+AB'+A'B-A'B'\le 2,
\label{Bell0}
\end{equation}
an inequality which is easily proved by just assuming the all the variables take the value +1 or -1.
 
It is now known that the inequality (\ref{Bell}) can be violated both by Nature and by quantum mechanics.

My explanation for this is as follows: The four possible outcomes with specified settings $A,A',B,B'$ constitute a set of four numbers that cannot all be observed by Alice nor by Bob during measurements and hence not by any observer. I will insist that such a set of variables is meaningless, in the sense that no theory made by humans should be able to describe this set of variables. This includes quantum mechanics and any potential theory that could be thought of to replace quantum mechanics. Any theory should concentrate on the (ideal) measurements made by Alice and Bob, each attached to one observer and to what is known to this observer: Her or his settings $a/a'$, respectively $b/b'$ and the resulting measurement, $x$ for Alice (ideal variable $\theta$) and $y$ for Bob (ideal variable $\psi$). 

I will generalize this to: For any allowable variable in any possible (quantum-like) theory one should be permitted to think of this variable as included in an experiment where the variable is exclusively attached to a particular observer. Only if the actual experiment is such that all real and imagined observers agree on some observation, this observation result may be seen as a property of the real world.

As is well known, several loophole-free Bell experiments have been performed during recent years. (For a discussion of loopholes; see Larsson [3]; the experiments are described in [4, 5].) To describe such experiments, it is in my opinion natural to use statistical language. As suggested by Richard Gill, I use the word `trial' for the choices of one setting each by Alice and Bob and the observation of one binary outcome by each, and I use the word `run' for a sequence of trials. During a run, settings are typically chosen anew, again and again, by some random mechanism. At the end of the run, Alice and Bob come together and then together have a large number $N$ of quadruples: setting Alice, outcome Alice, setting Bob, outcome Bob. Note that these are data, and must be modeled by some statistical model.

Look first at Alice's data and the model of these. The settings, although random, can also be thought of as chosen by Alice. Thus it is natural to have in the model a variable $z$, indicating the contribution to the setting by Alice. In statistical language this variable must be ancillary, that is, have a distribution that is independent of any parameter. By the conditionality principle of statistics, see for instance [1], every inference should be conditional, given $z$. Similarly, Bob has his model and his $z$ to be conditioned on. This gives two separate statistical inferences, one for Alice and one for Bob. The four estimates of mean outcomes: for Alice with settings $a$ and $a'$ and for Bob with settings $b$ and $b'$ are then the results of two different models and may well violate the CHSH inequality. Note that these are estimates from two separate sets of data, and there is no need to talk about any corresponding `real' vector consisting of 4 hypothetical perfect outcomes.

It is often said that the Bell theorem shows that quantum theory must violate either reality, locality or no-conspiracy. It may be inferred from these recent experiments that Nature also must violate either reality, locality or no-conspiracy. I do not want to drop locality, simply for the reason that I believe in relativity theory. In fact it seems that reality is violated in the sense that there exist certain `real' quantities that we are not allowed to talk about in any human-made theory. In the physical literature different concepts describing attitudes to the Bell experiment are introduced. For instance, in the language of Schmelzer [6], the attitude to rely on conditioning in the way that I have advocated, is related to superdeterminism, a concept which is further discussed in [7].

We must always hold open the possibility that a physical variable has to be connected to a particular observer. This is an important point of departure for me.This also implies that the quantum state of a given physical system may depend on the observer.

In this article, as in [1] and [8], a conceptual variable is defined as any variable attached to a person or by a group of communicating persons. A conceptual variable is called accessible if it, through an experiment or in other ways, is possible to assign numerical values to it; otherwise it is called inaccessible. For example, by Heisenberg's inequality the vector (position, momentum) for a particle is inaccessible. These notions are particularly interesting in the case of an epistemic process, a process to achieve knowledge. Epistemic processes are discussed in detail in [1]. Two basic such processes are statistical experiments and quantum measurements, but there are also other types of epistemic processes.

Going back to the Bell situation, the vector of four hypothetical ideal spin measurements is inaccessible, but in a very strong sense: The recent loophole-free experiments have indicated that even Nature does not permit any discussion of any consequences of regarding this vector as a realistic variable.

As will be thoroughly discussed below, the point of departure sketched here together with symmetry assumptions has implications both for the foundation for and the interpretation of quantum theory. First, it implies under certain conditions the Hilbert space formulation, including  introductions of operators for accessible variables and state vectors. Secondly it makes the following interpretation natural: State vectors, or more generally eigenvector spaces of operators, are in correspondence with focused questions to nature together with sharp answers to these questions. Arguments for this interpretation are discussed more fully in Section 7 below. For brevity, except for this introductory discussion of the Bell inequality, entanglement will not be further discussed in this paper.

Taking this loosely defined notion of conceptual variables as a basis, and then postponing the problem of making this notion concrete, allows us largely to stay agnostic concerning some of the debates on the interpretation of quantum theory and the debates between the different schools in statistical inference theory. However, the epistemic view in quantum theory turns out to be necessary, and an ideal observer may act as some kind of Bayesian in this theory.

The interpretation proposed here, is in fact similar to Herv\'{e} Zwirn's recent Convival Solipsism, see [9] and references there. In the latter paper, a more detailed discussion of the EPR paradox and the Bell inequalities is given. Comparisons with Everett's many world interpretations and variants of this are thoroughly discussed. Convival Solipsism allows quantum mechanics to be local even if it is at the price of reinterpreting the formalism in a rather radical way. I agree that such a reinterpretation has to be made. Zwirn argues that no satisfying solution can avoid mentioning the observer as playing a major role in the measurement. I completely agree with this.

To go a little more in detail, Convival Solipsism states that the physical world dynamics is entirely driven by the linear, deterministic dynamics of the Scr\"{o}dinger equation and nothing else. But it is assumed that our brain is not equipped to make us aware of all the subtleties of the superposed state, to perceive the universe in all its subtleties. In particular, Zwirn makes two assumptions: 1) The relative state assumption: Any state vector is relative to a given conscious observer and cannot be considered in an absolute way. 2) The hanging-on mechanism: A measurement is the awareness of a result by a concious observer whose consciousness selects at random (according to the Born rule) one branch of the entangled state vector written in the preferred basis and hangs-on to it.

Except for the point that I rather would have used the word `focusing' instead of `awareness', this interpretation seems to fit rather perfectly into my own views. Even so, I will in this article prefer to stay more agnostic regarding interpretations, except that I have a general epistemic view. If necessary, this view may be seen as a concrete specification of Rovelli's general Relational Quantum Mechanics [10] or of the more information-related views advocated for instance in [11].

Zwirn states also that the Born rule is not problematic by itself but the conditions of its use may be unclear. Again I agree. For my own derivation of the Born rule and other derivations, see Section 9 below.

In Sections 4-9 of this paper I concentrate on the derivation of the quantum formalism. It is crucial in this context that I assume that there are defined transformation groups on the spaces of conceptual variables that are discussed. This may be the group of all automorphisms on the space in which the conceptual variable varies, or it may be a concrete subgroup of this group.

In this paper I discuss such situations more closely with a focus on the more mathematical aspects of the notions and situations described above. I assume the existence of a concrete (physical) situation, and that there is a space $\Omega_\Phi$ of an inaccessible conceptual variable $\phi$  with a group $K$ acting on this space. There is an e-variable, an accessible conceptual variable, $\theta$ defined, a function on $\Omega_\Phi$. This $\theta$ varies on a space $\Omega_\Theta$, and the group $K$ may induce a transformation group $G$ on $\Omega_\Theta$. 

A very simple situation is when $\phi$ is a spin vector, and $\theta$ is a spin component in a given direction, but this paper focuses on more general situations. Also, in the simple spin situation the natural group $K$ for the spin vector does not directly induce groups on the components, so one must seek special solutions for this case; see a brief discussion in Section 6 below.

When there are several potential accessible variables, I will denote this by a superscript $a$: $\theta^a$ and $G^a =\{g^a\}$. Note that in Chapter 4 of [1] I used different notations for the groups: $G$ and $\tilde{G}^a$ for what is here called $K$ and $G^a$. The group $G^a$ which was refered to there, is here the subgroup of $K$ corresponding to $G^a$, denoted by $H^a$ below. Both here and in [1] I will use the word `group' as synonymous to `group action' or transformation group on some set, not as an abstract group.

The article has some overlap with the paper [8], but the perspective here is wider.

To avoid misunderstandings, I do not claim that my derivations should compete with the deep investigations recently on deriving the Hilbert space structure from physical assumptions [12, 13,14,15,16]. By using group representation theory, I assume in a sense some Hilbert space structure. As is stated in [17] though, there is a problem connecting the above general derivations to the many different interpretations of quantum theory. By contrast, the derivation presented here is tied to a particular interpretation: A general epistemic interpretation. This is also elaborated on in [1].

Group representation theory in discussing quantum foundation has also been used other places; see for instance [18]. In quantum field theory and particle physics theory, the use of group representation theory is crucial [19].

Section 2 gives some background. In Section 3 I introduce some basic group theory that is needed in the paper. Then in Sections 4-9 I formulate my approach to the foundation of quantum theory. The basis is conceptual variables with group actions defined on them. From this, the ordinary quantum formalism is developed, and it is shown how operators corresponding to accessible physical variables may be derived. The Born formula is derived by using three assumptions: 1) A focused likelihood principle, derived in [1] from the ordinary likelihood principle of statistics; 2) An assumption that the actual agent has ideals which can be modeled by a perfectly rational ideal agent; 3) The assumption that the initial state may be connected to a maximally accessible variable. In Section 10 various approaches to ordinary statistical inference are discussed, and in Section 11 an example is given which lies in the border area between quantum theory and Bayesian statistical inference. Section 12 discusses briefly quantum measurement theory, while a final discussion is given in Sections 13-15. 

\section{Some quantum theory and more discussions on interpretation}

It is assumed that the reader is familiar with the ordinary quantum theory for discrete observables: For each physical system there is a Hilbert space $\mathcal{H}$, and each observable $\theta$ on the system is associated with a self-adjoint operator $A$. The possible values of $\theta$ are the eigenvalues of $A$. The states of the system are the unit vectors $|u\rangle$ of $\mathcal{H}$, and when $|u\rangle$ is an eigenvector of $A$ corresponding to an eigenvalue $u_k$, then $\theta=u_k$ with certainty. If necessary, the minimum amount of theory is for instance given in [20].

The purpose of the paper [8] has been to rederive essential elements of this formal theory from assumptions about conceptual variables. On the following sections a part of the discussion of [8] is included for completeness.

As is well known, the probabilities of quantum theory are calculated by the Born rule. One version of this rule is: Assume that the physical system has been prepared in the state $|u\rangle$, corresponding to $\theta^a =u_k$ for some observable $\theta^a$, and let the aim be to measure another observable $\theta^b$. Then
\begin{equation}
P(\theta^b =v_j )=\langle u |\Pi_j | u\rangle,
\label{Born}
\end{equation}
where $\Pi_j$ is the projector upon the eigenspace corresponding to the eigenvalue $\theta^b =v_j$. The Born rule is derived from a reasonable set of assumptions in [1]; see a discussion in Section 9 below.

The reader is also probably aware of the fact that there exist several interpretations of quantum theory, and that the discussions between the supporters of the different interpretations is still going on. During the recent years there has been held a long range of international conferences on the foundation of quantum mechanics. A great number of interpretations have been proposed; some of them look very peculiar to the laymen. For instance, the many worlds interpretation assumes that there exist millions or billions of parallel worlds, and that a new world appears every time when one performs a measurement.

On two of these conferences recently there was taken an opinion poll among the participants [21, 22]. It turned out to be an astonishing disagreement on many fundamental and fairly simple questions. One of these questions was: Is the quantum mechanics a description of the objective world, or is it only a description of how we obtain  knowledge about reality? The first of these descriptions is called ontological, the second epistemic. Up to now most physicists have supported the ontological or realistic interpretation of quantum mechanics, but versions of the epistemic interpretation have received a fresh impetus during the recent years.

I look upon my book 'Epistemic Processes' [1] as a contribution to this debate. An epistemic process can denote any process to achieve knowledge. It can be a statistical investigation or a quantum mechanical measurement, but it can also be a simpler process. The book starts with an informal interpretation of quantum states, which in the traditional theory has a very abstract definition. In my opinion, a quantum state can under wide circumstances be connected to a focused question and a sharp answer to this question.

A related interpretation is QBism, or quantum Bayesianism, see Fuchs [23, 24, 25] and von Baeyer [26]. The predictions of quantum mechanics involve probabilities, and a QBist interpret these as purely subjective probabilities, attached to a concrete observer.  Many elements in QBism represent something completely new in relation to classical physical theory, in relation to many people's conception about science in general and also to earlier interpretations of quantum mechanics. The essential thing is that the observer plays a role that can not be eliminated. The single person's comprehension of reality can differ from person to person, at least at a given point of time, and this is in principle all that can be said.

Such an understanding can in my opinion be made valid for very many aspects of reality. We humans can have a tendency to experience reality differently. Partly, this can be explained by the fact that we give different meaning to the concepts we use. Or we can have different contexts for our choices. An important aspect is that we focus differently.

Herv\'{e} Zwirn's views on QBism, which I largely agree with, are given in [27].

The statement that the state of a physical system may be connected an observer (or a group of communicating observers), is discussed in the Introduction above.

By using group theory and group representation theory, I aim at studying a situation involving conceptual variables mathematically, and it seems to appear that an essential part of the quantum formulation can be derived under very weak conditions. My opinion is that what we can develop from such results can be crucial for our views on the world around us. Empirically, the quantum formalism has turned out to give a very extensive description of our world as we know it, in physical situations in microcosmos an all-embracing description. 

In decision situations and in cognitive modeling it has also been fruitful to look at a quantum description, see [28, 29]. In a decision situation the decision variable may be so extensive that it is impossible for the person in question to make a decision; this variable is then called inaccessible. The person can then focus on a simpler, accessible, decision variable, in such a way that it is possible to make a partial decision. 

Sometimes it can be necessary to have an epistemic way to relate to the world; it can simply be useful to seek knowledge. We can get knowledge on certain issues by focusing on certain questions to the world around us, and our knowledge depends on the answers we obtain to these questions. And this is all we can achieve.

Following such an opinion, it can be argued that for certain phenomena there exists no other state concept than this (subjective) attached to each single person. This statement must be made precise to be understood in the correct way. First, it is connected to an ideal observer. Secondly, groups of observers that communicate, can go in and act as one observer when a concrete measurement is focused on. When all potential observers agree on the answer to a measurement, there is a strong indication that this measurement represents an objective property of reality. Thus the objective world exists; it is the state attached to certain aspects of the world that in certain cases must be connected to an observer (or to several communicating observers). 

Nevertheless, these are aspects of physics – and science – which can be surprising for many people, but in my opinion such viewpoints may be necessary, not only in physics, but also in many other contexts. 

Here is one remark concerning QBism, which can be said to represent a variant of this view: Subjective Bayes-probabilities have also been in fashion among groups of statisticians. In my opinion it can be very fruitful to look for analogies between statistical inference theory and quantum mechanics, but then one must look more broadly upon statistics and statistical inference theory, not only focus on subjective Bayesianism. This is only one of several philosophies that can form a basis for statistics as a science. Studying connections between these philosophies, is an active research area today; some results are given in Section 10 below. From this discussion one might infer that an interesting version of Bayesianism is an objective Bayesianism, with a prior based on group actions.

\section{Variables and group actions}

Let $\phi$ be an inaccessible conceptual variable varying in a space $\Omega_\Phi$. It is a basic philosophy of the present paper that I always regard groups as group actions or transformations, acting on some space.

Starting with $\Omega_\Phi$ and a group $K$ acting on $\Omega_\Phi$, let $\theta(\cdot)$ be an accessible function on $\Omega_\Phi$, and let $\Omega_\Theta$ be the range of this function.

As mentioned in the Introduction, I regard `accessible' and `inaccessible' as primitive notions. $\Omega_\Theta$ and $\Omega_\Phi$ are equipped with topologies, and all functions are assumed to be Borel-measurable.
\bigskip

 \textbf{Definition 1. }
\textit{The variable $\theta$ is maximally accessible if the following holds: If $\theta$ can be written as $\theta=f(\psi)$ for a function $f$ that is not bijective, the conceptual variable $\psi$ is not accessible. In other words: $\theta$ is maximal under the partial ordering defined by $\alpha\le \beta$ iff $\alpha=f(\beta )$ for some function $f$.}
\bigskip

Note that this partial ordering is consistent with accessibility: If $\beta$ is accessible and $\alpha=f(\beta )$, then $\alpha$ is accessible. Also, $\phi$ is an upper bound under this partial ordering. The existence of maximally accessible conceptual variables follows then from Zorn's lemma.
\bigskip

\textbf{Definition 2.}
\textit{The accessible variable $\theta$ is called\emph{ permissible} if the following holds: $\theta(\phi_1)=\theta(\phi_2)$ implies $\theta(k\phi_1 )=\theta(k\phi_2 )$ for all $k\in K$.}
\bigskip

With respect to parameters and subparameters along with their estimation, the concept of permissibility is discussed in some details in Chapter 3 in [30]. The main conclusion, which also is valid in this setting, is that under the assumption of permissibility one can define a group $G$ of actions on $\Omega_\Theta$ such that

\begin{equation}
(g\theta)(\phi):=\theta(k\phi);\ k\in K.
\label{kg}
\end{equation}

Herein I use different notations for the group actions $g$ on $\Omega_\Theta$ and the group actions $k$ on $\Omega_\Phi$; in contrast,  the same symbol $g$ was used in [30]. The background for that is
\bigskip

\textbf{Lemma 1.}
\textit{Assume that $\theta$ is a permissible variable. The function from $K$ to $G$ defined by (\ref{kg}) is then a group homomorphism.}
\bigskip

\textit{Proof.} Let $k_i$ be mapped upon $g_i$ by (\ref{kg}) for $i=1,2$. Then, for all $\phi\in \Omega_\Phi$ we have $(g_i\theta )(\phi)=\theta(k_i\phi )$. Assume that $k_2\phi $ is mapped to $\theta'=\theta(k_2 \phi )=( g_2\theta)(\phi)$. Then also $\theta(k_1 k_2 \phi  )=(g_1\theta)(k_2\phi )=(g\theta)(\phi)$ for some $g$. Thus $(g_1(g_2\theta))(\phi)=(g\theta)(\phi)$ for all $\phi$, and since the mapping is permissible, we must have $g=g_1g_2$.
\bigskip

It is important to define left and right invariant measures, both on the groups and on the spaces of conceptual variables. In the mathematical literature, see for instance [31, 32], Haar measures on the groups are defined (assuming locally compact groups). Right ($\mu_G$) and left ($\nu_G$) Haar measures on the group $G$ satisfy
\begin{eqnarray*}
\mu_G(Dg)=\mu_G(D), \ \mathrm{and}\ \nu_G(gD)=\nu_K(D)\\ \mathrm{for}\ g\in G\ \mathrm{and}\ D\subset G,\ \mathrm{respectively}.
\end{eqnarray*}

Next define the corresponding measures on $\Omega_\Theta$. As is commonly done, I assume that the group operations $(g_1,g_2)\mapsto g_1g_2$, $(g_1,g_2)\mapsto g_2g_1$ and $g\mapsto g^{-1}$ are continuous. Furthermore, I will assume that the action $(g,\theta)\mapsto g\theta$ is continuous. 

As discussed in Wijsman [33], an additional condition is that every inverse image of compact sets under the function $(g,\theta)\mapsto (g\theta,\theta)$ should be compact. A continuous action by a group $G$ on a space $\Omega_\Theta$ satisfying this condition is called \emph{proper}. This technical condition turns out to have useful properties and is assumed throughout this paper. When the group action is proper, the orbits of the group can be proved to be closed sets relative to the topology of $\Omega_\Theta$.

The following result, originally due to Weil is proved in [33]; for more details on the right-invariant case, see also [30].
\bigskip

\textbf{Theorem 1.}
\textit{The left-invariant measure measure $\nu$ on $\Omega_\Theta$ exists if the action of $G$ on $\Omega_\Theta$ is proper and the group is locally compact.}
\bigskip

The connection between $\nu_G$ defined on $G$ and the corresponding left invariant measure $\nu$ defined on $\Omega_\Theta$ is relatively simple: If for some fixed value $\theta_0$ of the conceptual variable the function $\beta$ on $G$ is defined by $\beta: g\mapsto g\theta_0$, then $\nu(E)=\nu_G (\beta^{-1}(E))$.This connection between $\nu_G$ and $\nu$ can also be written $\nu_G(dg)=d\nu(g\theta_0))$, so that $d\nu(hg\phi_0)=d\nu (g\phi_0)$ for all $h, g \in G$.

Note that $\nu$ can be seen as an induced measure on each orbit of $G$ on $\Omega_\Theta$, and it can be arbitrarily normalized on each orbit. $\nu$ is finite on a given orbit if and only if the orbit is compact. In particular, $\nu$ can be defined as a probability measure on $\Omega_\Theta$ if and only if all orbits of $\Omega_\Theta$ are compact. Furthermore, $\nu$ is unique only if the group action is transitive.

In a corresponding fashion, a right invariant measure can be defined on $\Omega_\Theta$. This measure satisfies $d\mu (gh\phi_0)=d\mu (g\phi_0)$ for all $g,h\in G$. In many cases the left invariant measure and the right invariant measure are equal.

\section{Operators and quantization}

In the quantum-mechanical context defined in [1], $\theta$ is an accessible variable, and one should be able to introduce an operator associated with $\theta$. The following discussion which is partly inspired by [34]. considers an irreducible unitary representation of $G$ on a complex Hilbert space $\mathcal{H}$.

\subsection{A brief discussion of group representation theory}
 
A group representation of $G$ is a continuous homomorphism from $G$ to the group of invertible linear operators $V$ on some vector space $\mathcal{H}$:
\begin{equation}
V(g_1 g_2 )=V(g_1 )V(g_2 ).
\label{3}
\end{equation}
It is also required that $V(e)=I$, where $I$ is the identity, and $e$ is the unit element of $G$. This assures that the inverse exists: $V(g)^{-1}=V(g^{-1})$. The representation is unitary if the 
operators are unitary ($V(g)^{\dagger}V(g)=I$). If the vector space is finite-dimensional, we have a representation $D(V)$ on the square, invertible matrices. For any 
representation $V$ and any fixed invertible operator $U$ on the vector space, we can define a new equivalent  representation as $W(g)=UV(g)U^{-1}$. One can prove that two 
equivalent unitary representations are unitarily equivalent; thus $U$ can be chosen as a unitary operator.

A subspace $\mathcal{H}_1$ of $\mathcal{H}$ is called invariant with respect to the representation $V$ if $u\in \mathcal{H}_1$ implies $V(g)u\in \mathcal{H}_1$ for all $g\in G$. The null-space $\{0\}$ and the whole space
$\mathcal{H}$ are trivially invariant; other invariant subspaces are called proper. A group representation $V$ of a group $G$ in $\mathcal{H}$ is called irreducible if it has no proper invariant subspace.
A representation is said to be fully reducible if it can be expressed as a direct sum of irreducible subrepresentations. A finite-dimensional unitary representation of any group 
is fully reducible. In terms of a matrix representation, this means that we can always find a  $W(g)=UV(g)U^{-1}$ such that $D(W)$ is of minimal block diagonal form. Each one of 
these blocks represents an irreducible representation, and they are all one-dimensional if and only if $G$ is Abelian. The blocks may be seen as operators on subspaces of the 
original vector space, i.e., the irreducible subspaces. The blocks are important in studying the structure of the group.

A useful result is Schur's Lemma:
\bigskip

\textit{Let $V_1$ and $V_2$ be two irreducible representations of a group $G$; $V_1$ on the space $\mathcal{H}_1$ and $V_2$ on the space $\mathcal{H}_2$. Suppose that there exists a linear map $T$ from $\mathcal{H}_1$ to 
$\mathcal{H}_2$ such that}
\begin{equation}
V_2 (g)T(v)=T(V_1 (g)v)
\label{4}
\end{equation}
\textit{for all $g\in G$ and $v\in\mathcal{H}_1$.}

\textit{Then either $T$ is zero or it is a linear isomorphism. Furthermore, if $\mathcal{H}_1=\mathcal{H}_2$, then $T=\lambda I$ for some complex number $\lambda$.}
\bigskip

Let $\nu$ be the left-invariant measure of the space $\Omega_\Theta$ induced by the group $G$, and consider in this connection the Hilbert space $\mathcal{H}=L^2 (\Omega_\Theta ,\nu)$. Then the left-regular 
representation of $G$ on $\mathcal{H}$ is defined by $U^{L}(g)f(\phi)=f(g^{-1}\phi)$. This representation always exists, and it can be shown to be unitary, see [34].

If $V$ is an arbitrary representation of a compact group $G$ in some Hilbert space $\mathcal{H}$, then there exists in $\mathcal{H}$ a new scalar product defining a norm equivalent to the initial one, relative to which $V$ 
is a unitary representation of $G$.

For references to some of the vast literature on group representation theory, see Appendix A.2.4 in [30].

\subsection{A resolution of the identity}
\label{sec:3.2}

In the following I assume that the group $G$ has representations that give square-integrable coherent state systems (see page 43 of [34]). For instance this is the case for all representations of compact semisimple groups, representations of discrete series for real semisimple groups, and some representations of solvable Lie groups.

Let $G$ be an arbitrary such group, and let $V(g)$ be one of its unitary  irreducible representations acting on a Hilbert space $\mathcal{H}$. Assume that $G$ is acting transitively on a space $\Omega_\Theta$, and fix $\theta_0\in\Omega_\Theta$. Then every $\theta\in\Omega_\Theta$ can be written as $\theta=g\theta_0 $ for some $g\in G$. I also assume that the isotropy group of $G$ is trivial. Then this establishes a one-to-one correspondence between $G$ and $\Omega_\Theta$. 

Also, fix a vector $|\theta_0\rangle\in\mathcal{H}$, and define the coherent states $|\theta\rangle=|\theta(g)\rangle=V(g)|\theta_0\rangle$. With $\nu$ being the left invariant measure on $\Omega_\Theta$, introduce the operator
\begin{equation}
T=\int |\theta(g)\rangle\langle\theta(g)|d\nu(g\theta_0).
\label{5}
\end{equation}
Note that the measure here is over $\Omega_\Theta$, but the elements are parametrized by $G$. $T$ is assumed to be a finite operator.
\bigskip

\textbf{Lemma 2.}
\textit{$T$ commutes with every $V(h); h\in G$.} 
\bigskip

\textit{Proof.} $\ \ V(h)T=$
\begin{eqnarray*}
\int V(h) |\theta(g)\rangle\langle\theta(g)|d\nu(g\theta_0)
=\int |\theta(hg)\rangle\langle\theta(g)|d\nu(g\theta_0)\\=\int |\theta(r)\rangle\langle\theta(h^{-1}r)|d\nu(h^{-1}r\theta_0 ).
\end{eqnarray*}
Since $|\theta(h^{-1}r)\rangle=V(h^{-1}r)|\theta_0\rangle =V(h^{-1})V(r)|\theta_0\rangle =V(h)^\dagger |\theta(r)\rangle$, we have $\langle \theta(h^{-1}r)|=\langle\theta(r)|V(h)$, and since the measure $\nu$ is left-invariant, it follows that $V(h)T=TV(h)$. $\Box$
\bigskip

From the above and Schur's Lemma it follows that $T=\lambda I$ for some $\lambda$. Since $T$ by construction only can have postive eigenvalues, we must have $\lambda >0$. Defining the measure $d\mu(\theta)=\lambda^{-1}d\nu(\theta)$ we therefore have the important resolution of the identity
\begin{equation}
\int |\theta\rangle\langle\theta |d\mu(\theta)=I.
\label{6}
\end{equation}
For a more elaborate similar construction taking into account the socalled isotropy subgroup, see Chapter 2 of [34]. In [8] a corresponding resolution of the identity is derived for states defined through representations of the group $K$ acting on $\Omega_\Phi$. 

\subsection{Quantum operators}

Let now $\theta$ be a maximally accessible variable, and let $G$ be a group acting on $\theta$.

In general, an operator corresponding to $\theta$ may be defined by
\begin{eqnarray}
A=A^\theta
=\int \theta |\theta\rangle\langle\theta | d\mu (\theta).
\label{7}
\end{eqnarray}
$A$ is defined on a domain $D(A)$ of vectors $|v\rangle\in\mathcal{H}$ where the integral defining $\langle v|A|v\rangle$ converges.

This mapping from an accessible variable $\theta$ to an operator $A$ has the following properties:

(i) If $\theta=1$, then $A=I$.

(ii) If $\theta$ is real-valued, then $A$ is symmetric (for a definition of this concept for operators and its relationship to self-adjointness, see [36].)

(iii) The change of basis through a unitary transformation is straightforward.

For further important properties, we need some more theory. First consider the situation where we regard the group $G$ as generated by a group $K$ defined on the space of an inaccessible variable $\phi$. This represents no problem if the mapping from $\phi$ to $\theta$ is permissible, a case discussed in [8], and in this case the operators corresponding to several accessible variables can be defined on the same Hilbert space. In the opposite case we have the following theorem.
\bigskip

\textbf{Theorem 2.} \textit{Let $H$ be the subgroup of $K$ consisting of any transformation $h$ such that $\theta(h\phi)=g\theta(\phi)$ for some $g\in G$. Then $H$ is the maximal group under which the e-variable $\theta$ is permissible.}
\bigskip

\textit{Proof.} Let $\theta(\phi_1) =\theta(\phi_2)$ for all $\theta\in\Theta$. Then for $h\in H$ we have $\theta(h\phi_1)=g\theta(\phi_1)=g\theta(\phi_2)=\theta (h\phi_2)$, thus $\theta$ is permissible under the group $H$. For a larger group, this argument does not hold. $\Box$
\bigskip

Next look at the mapping from $\theta$ to $A^\theta$ defined by (\ref{7}). 
\bigskip

\textbf{Theorem 3.} \textit{ For $g\in G$, $V(g^{-1})AV(g)$ is mapped by $\theta'=g\theta$.}
\bigskip

\textit{Proof.} $V(g^{-1})AV(g)=$
\begin{eqnarray*}
\int \theta |g^{-1}\theta\rangle\langle g^{-1}\theta |d\mu(\theta)=\int g\theta |\theta\rangle\langle \theta |d\mu(g\theta).
\end{eqnarray*}
Use the invariance of $\mu$. $\Box$
\bigskip

Further properties of the mapping from $\theta$ to $A$ may be developed in a similar way.
The mapping corresponds to the usual way that the operators are allocated to observables in the quantum mechanical literature. But note that this mapping comes naturally here from the notions of conceptual variable and e-variables on which group actions are defined.

\subsection{The spectral theorem and operators for functions of $\theta$}

Assume that $\theta$ is real-valued. Then based on the spectral theorem (e.g., [36]) we have that there exists a projectionvalued measure, $E$ on $\Omega_\Theta$ such that for $|v\rangle\in D(A)$
\begin{equation}
\langle v|A|v\rangle =\int_{\sigma(A)} \theta d\langle v|E(\theta)|v\rangle .
\label{8}
\end{equation}
Here $\sigma(A)$ is the spectrum of $A$ as defined in [36].
The case with a discrete spectrum is discussed in the next subsection.

A more informal way to write (\ref{8}) is

\[A=\int_{\sigma(A)}\theta dE(\theta).\]

This defines a resolution of the identity
\begin{equation}
\int_{\sigma(A)}dE(\theta )=I.
\label{9}
\end{equation}

From this, we can define the operator of an arbitrary Borel-measurable function of $\theta$ by
\begin{equation}
A^{f(\theta)}=\int_{\sigma(A)} f(\theta) dE (\theta).
\label{10}
\end{equation}

Important special cases include $f(\theta)=I(\theta\in B)$ for sets $B$. Another important observation is the following: Any accessible variable can be written as $f(\theta)$, where $\theta$ is maximally accessible.

A further important case is connected to statistical inference theory in the way it is advocated in [1]. Assume that there are data $X$ and a statistical model for these data of the form $P(X\in C|\theta)$ for sets $C$. Then a positive operator-valued measure (POVM) on the data space can be defined by
\begin{equation}
M(C)=\int_{\sigma(A)} P(X\in C |\theta )dE(\theta).
\label{11}
\end{equation}
The density of $M$ at a point $x$ is called the likelihood effect in [1], and is the basis for the focused likelihood principle formulated there; see Section 8 below.

Finally, given a probability measure with density $\pi(\theta)$ over the values of $\theta$, one can define a density operator $\sigma$ by
\begin{equation}
\sigma=\int_{\sigma(A)} \pi(\theta)dE (\theta).
\label{12}
\end{equation}

In [1] the probability measure $\pi$ was assumed to have one out of three possible interpretations: 1) as a Bayesian prior,  2) as a Bayesian posterior or 3) as a frequentist confidence distribution (see [11]).

\subsection{The case of a purely discrete spectrum}

Note that the construction in the previous subsections can be made for any accessible conceptual variable $\theta$. Assume now that $A$ has a purely discrete spectrum. $\theta$ may be a vector. Then $A$ is a tensor product of commuting operators. 

Let the eigenvalues be $\{u_j\}$ and let the corresponding eigenspaces be $\{V_j\}$. The vectors of these eigenspaces are defined as quantum states, and as in [1], each eigenspace $V_j$ is associated with a question `What is the value of $\theta$?' together with a definite answer `$\theta =u_j$'. This assumes that the set of values of $\theta$ can be reduced to this set of eigenvalues, which I will  justify as follows.

\bigskip

\textbf{Theorem 4.}
\textit{ Assume that the group $G$ is transitive on $\Omega_\Theta$. Then if one of the values of $\theta$ is an eigenvalue of $A$, all the values of $\theta$ are eigenvalues of $A$, and $G$ is a permutation group on these eigenvalues.}
\bigskip

\textit{Proof.} For each $j$, let $|j\rangle$ be an eigenvector of $A$ with eigenvalue $u_j$, and let $g\in G$.  By Theorem 3 we have that the operator $V(g^{-1})AV(g)$ is mapped by $g\theta$. Fix $\theta_0\in \Omega_\Theta$, and assume that $\theta_0 =u_j$ for some $j$. We need to show that $g\theta_0$ is another eigenvalue for $A$, which follows from the fact that $|V(g^{-1})AV(g)-\lambda I|=|A-\lambda I|$, so that these two determinants have the same zeros. It is assumed that $G$ is transitive, so this gives all values in $\Omega_\Theta$ and at the same time all eigenvalues of $A$. $\Box$
\bigskip

It is crucial for this whole discussion that the group $G$ is transitive on $\Omega_\Theta$. If this is not the case, one can use model reduction. The following principle has been proved useful in statistical inference. (See discussion and examples in [1], and an important application in [37, 38].)
\bigskip

\textbf{Principle}
\textit{When a group $G$ is defined on the parameter space, every model reduction should be to an orbit or to a set of orbits of this group.}
\bigskip

In a quantum setting every model should be reduced to a single orbit. \emph{This can be seen as a general principle for quantization.}
\bigskip

We also have the following:
\bigskip

\textbf{Theorem 5.} \textit{Assume that the set of measurable values of $\theta$ is restricted to the above eigenvalues. Then $\theta$ is maximally accessible if and only if each eigenspace $V_j$ is one-dimensional.}
\bigskip

\textit{Proof.} The assertion that there exists an eigenspace that is not one-dimensional, is equivalent with the following: Some eigenvalue $u_j$ correspond to at least two orthogonal eigenvectors $|j\rangle$ and $|i\rangle$. Based on the spectral theorem, the operator $A$ corresponding to $\theta$ can be written as $\sum_r u_r P_r$, where $P_r$ is the projection upon the eigenspace $V_r$. Now define a new e-variable $\psi$ whose operator $B$ has the following properties: If $r\ne j$, the eigenvalues and eigenspaces of $B$ are equal to those of $A$. If $r=j$, $B$ has two different eigenvalues on the two one-dimensional spaces spanned by $|j\rangle$ and $|i\rangle$, respectively, otherwise its eventual eigenvalues are equal to $u_j$ in the space $V_j$. Then $\theta=\theta(\psi)$, and $\psi\ne\theta$ is inaccessible if and only if $\theta$ is maximally accessible. This construction is impossible if and only if all eigenspaces are one-dimensional. $\Box$

\section{Coupling different foci together}

\subsection{Two maximally accessible variables}

Let $\lambda^1$ and $\lambda^2$ be two maximally accessible variables, and let $\phi=(\lambda^1,\lambda^2)$. An example may be $\lambda^1=$position and $\lambda^2$=momentum of a particle. Assume that a group $G^1$ is defined on $\Omega_{\lambda^1}$, a group $G^2$ on $\Omega_{\lambda^2}$, and let $K=G^1\otimes G^2$. In this case it is easy to see that each $\lambda^i (\phi)$ is a permissible function of $\phi$, and the discussion of subsection 4.2 leads to resolutions of the identity
\begin{equation}
\int |\lambda^i \rangle\langle\lambda^i |d\mu^i (\lambda^i)=I.
\label{M1}
\end{equation}

It may be convenient to treat each $\lambda^i$ explicitly as a function of $\phi$, and let $\mu^i$ be the marginalization of a measure $\tau^i$ on $\Omega_\phi$. Then it follows that
\begin{equation}
\int |\lambda^i (\phi)\rangle\langle\lambda^i (\phi)|d\tau^i (\phi)=I,
\label{M2}
\end{equation}
and we may define operators
\begin{equation}
A^i =\int \lambda^i (\phi) |\lambda^i (\phi)\rangle\langle\lambda^i (\phi)|d\tau^i (\phi).
\label{M3}
\end{equation}

One point is that both operators can be defined on the same Hilbert space. By maximality, only one of $\lambda^1$ or $\lambda^2$ can be measured on the system at a given time. This is a manifestation of Niels Bohr's complementarity.

I will continue to discuss the case where the operator corresponding to $\lambda$ has a discrete spectrum.
 
 \subsection{Maximally accessible discrete variables}
 
In this section consider a Hilbert space $\mathcal{H}$ of finite dimension $d$. Again let $\phi$ be an inaccessible conceptual variable. For an index set $\mathcal{A}$, focus on $\lambda^a$ for $a\in\mathcal{A}$, a set of maximally accessible variables. Assume now that each mapping $\phi\mapsto\lambda^a (\phi)$ is permissible, so that $K$ induces a group $G^a$ acting upon $\lambda^a$. Then each $\lambda^a$ corresponds to a unique operator $A^a$, and this operator has the spectral decomposition
\begin{equation}
A^a=\sum_j u_j^a |a;j\rangle\langle a;j|.
\label{13}
\end{equation}

Again by maximality, only one $\lambda^a$ can be measured on the system at a given time.

\subsection{Non maximally accessible discrete variables}

First go back to the maximal symmetrical epistemic setting. Let again $\lambda^a=\lambda^a(\phi)$ be as in the previous subsection. Let $t^a$ be an arbitrary function on the range of $\lambda^a$, and let us focus on $\theta^a =t^a(\lambda^a)$ for each $a\in\mathcal{A}$.

 Let the Hilbert space be as in the previous subsection, and suppose that it has an orthonormal basis that can be written in the form $|a;i\rangle$ for $i=1,...,d$. Let $\{u_i^a\}$ be the values of $\lambda^a$, and let $\{s_j^a\}$ be the values of $\theta^a$. Define $C_j^a=\{i:t^a(u_i^a)=s_j^a\}$, and let $V_j^a$ be the space spanned by $\{|a;i\rangle :i\in C_j^a\}$. Let $\Pi_j^a$ be the projection upon $V_j^a$.
 
 Then we have the following interpretation of any $|a;i\rangle\in V_j^a$. (1) the question: `What is the value of $\theta^a$?' has been posed, and (2) we have obtained the answer $\theta^a=s_j^a$. Note that in this case, several pairs $(a,i)$ correspond to a given vector $|v\rangle$.
 
 From the above construction we may also define the operator connected to the e-variable $\theta^a$ as
 \begin{equation}
 A^a=\sum_j s_j^a \Pi_j^a =\sum_i t^a(u_i^a)|a;i\rangle\langle a;i|.
 \label{14}
\end{equation}
 Note that this gives all possible states and all possible values corresponding to the accessible e-variable $\theta^a$. Unless the function $t^a$ is bijective, the operator $A^a$ has no longer distinct eigenvalues. 
 
\section{The spin case}

Let $\phi$ be the total spin vector of a particle, and let $\theta^a =\|\phi\| \mathrm{cos}(\phi ,a)$ be the component of $\phi$ in the direction $a$. If we have a coordinate system, the components $\theta^x , \theta^y$ and $\theta^z$ are of special interest. As pointed out by Yoh Tanimoto, the following example shows that these components are not permissible if $K$ is the rotation group for some fixed $\|\phi\|$:

Take $\phi_1 =(1,0,0)$, $\phi_2 = (0,1,0)$, and let $k$ be the rotation in the $xz$-plane such that $k\phi_1 =(0,0,1)$ and $k\phi_2 =(0,1,0)$. Then the $z$-components of $\phi_1$ and $\phi_2$ are equal, but the $z$-components of $k\phi_1$ and $k\phi_2$ are different.

But note that (contrary to the discussion in [8]) the homomorphism from the group $K$ acting on $\phi$ to the group $G^a$ acting on $\theta^a$ (here the permutation group) was not used in Section 4 above. If necessary, given a spin component $\theta^a$ in the direction $a$, one can consider the largest subgroup for which the mapping from $\phi$ to $\theta^a (\phi)$ is permissible. Such a subgroup always exists (Theorem 2 above). In the present case it consists of all  rotations around $a$ together with a reflection in the plane orthogonal to $a$.

The orbits of the corresponding group acting on $\theta^a$ are simply $\pm c$ for $c>0$. By using a suitable scale and the principle that a reduced model should be to an orbit or a set of orbits of the relevant group, one can argue that the possible values of $\theta^a$ should be $-r, -r+1,..., r-1, r$ for a non-negative integer or half-integer $r$.

The purpose of the general discussion in [8] was to give a firm basis to the statement that a common Hilbert space can be chosen to find operators for each variable $\theta^a$. For spin components thus a special argument must be given for this. 

Spin coherent states and their connection to the group SU(2) are discussed thoroughly in [34] and [39]. I will follow parts of [34] without going into details. It is crucial that any irreducible representation $D$ is given by a nonnegative integer or half-integer $r$: $D(k)=D^r (k)$, $\mathrm{dim}(D^r )=d=2r+1$. In the representation space $\mathcal{H}=\mathcal{H}_r$ the canonical basis $|r;m\rangle$ exists, where $m$ runs from $-r$ to $r$ in unit steps. 
Seen as matrices, the operators $D(k)$ may be taken to act on the $d$-dimensional vector space $\mathbbm{C}^d$. These matrix representation is discussed in many books, for instance [40]. It is stated on op. cit., page 129 that $D$ is a single-valued representation of the group SO(3) when $r$ is an integer, a double-valued representation when $r$ is half of an odd integer. In all cases it is a representation of SU(2).

The infinitesimal operators $A^{\pm}=A^x \pm i A^y$, $A^0 =A^z$ of the group representation $D^r$ satisfy the commutation relations

\begin{equation}
[A^0 , A^{\pm}]=\pm A^{\pm},\ \ [A^- ,A^+ ]=-2 A^0 .
\label{commutation}
\end{equation}
The operators  $A^x$, $A^y$ and $A^z$ are related to infinitesimal rotations around the $x$-axis, $y$-axis and $z$-axis, respectively, and we take $\bm{A}=(A^x ,A^y ,A^z )$. The representation space vectors $|r;m\rangle$ are eigenvectors for the operators $A^0$ and $\bm{A}^2 = (A^x)^2 +(A^y)^2 + (A^z)^2$:

\begin{equation}
A^0 |r;m\rangle =m|r; m\rangle ,\ \ \bm{A}^2 |r;m\rangle =r(r+1)|r;m\rangle .
\label{eigen}
\end{equation} 

The operator $\mathrm{exp}[i\omega(\bm{n}\cdot\bm{A})]$, $\|\bm{n}\| =1$, describes the rotation by the angle $\omega$ around the axis directed along $\bm{n}$. In [34] this was used to describe the coherent states $D(k)|\phi_0\rangle$ in various ways. The ket vector $|\phi_0\rangle$ may be taken as $|r;m\rangle$ for a fixed $m$; the simplest choice is $m=-r$. 

\section{Interpretation of quantum states and variables}

The following simple observation should be noted, and is in correspondence with the ordinary textbook interpretation of quantum states: Trivially, every vector $|v\rangle$ is the eigenvector of \emph{some} operators. Assume that there is one such operator $A$ that is physically meaningful, and for which $|v\rangle$ is also a non-degenerate eigenvector, say with a corresponding eigenvalue $u$. Let $\lambda$ be a physical variable associated with $A$. Then $|v\rangle$ can be interpreted as the question `What is the value of $\lambda$?' along with the definite answer $\lambda=u$.

More generally, accepting operators with non-gerenerate eigenspaces (corresponding to observables that are accessible, but not maximally accessible), each eigenspace can be interpreted as a question along with an answer to this question.

Binding together these two paragraphs, we can think of the case where $\lambda$ is a vector, such that each component $\lambda_i$ corresponds to an operator $A_i$, and these operators are mutually commuting. Then $A=\bigotimes_i A_i$ has eigenspaces which can be interpreted as a set of questions `What is the value of $\lambda_i\ i=1,2,...$?' together with sharp answers to these questions. In the special case of systems of qubits, H\"{o}hn and Wever [67] have recently proved that there is a one-to-one correspondence here.

The following is proven in [1, 41] under certain general technical conditions, and also specifically in the case of spin/ angular momentum: Given a vector $|v\rangle$ in a Hilbert space $\mathcal{H}$ and a number $u$, there is at most one pair $(a,j)$ such that $|a;j\rangle=|v\rangle$ modulus a phase factor, and $|a;j\rangle$ is an eigenvector of an operator $A^a$ with eigenvalue $u$. 

The main interpretation in [1, 41] is motivated as follows: Suppose the existence of such a vector $|v\rangle$ with $|v\rangle=|a;j\rangle$ for some $a$ and $j$. Then the fact that the state of the system is $|v\rangle$ means that one has focused on a question (`What is the value of $\lambda^a$?') and obtained the definite answer ($\lambda^a=u$.) The question can be associated with the orthonormal basis  $\{|a;j\rangle ;j=1,2,...,d\}$, equivalently with a resolution of the identity $I=\sum_j |a;j\rangle\langle a;j|$.

The general technical result of [1, 41] is also valid in the case where $\lambda^a$ and $u$ are real-valued vectors.
After this we are left with the problem of determining the exact conditions under which \emph{all} vectors $|v\rangle\in \mathcal{H}$ in the non-degerate discrete case and all projection operators in the general case can be interpreted as above. This will require a rich index set $\mathcal{A}$ determining the index $a$. This problem will not be considered further here, but this is stated as a general question to the quantum community in [41]. But from the evidence above, I will in this paper rely on the conjecture that each quantum state/ eigenvector space can be associated in a unique way with a question-and-answer pair. 

When $\lambda$ is a continuous variable or even a more general variable, we can still interpret the eigenspaces of the operator $A$ as questions `What is the value of $\lambda$?' together with answers in terms of intervals for $\lambda$. This is related to the spectral decomposition of $A$, which gives the resolution of the identity (recall (\ref{9}))
\begin{equation}
I=\int_{\sigma(A)} dE(\lambda).
\label{111}
\end{equation}

This resolution of the identity is tightly coupled to the question `What is the value of $\lambda?$', and it implies projections related to indicators of intervals/sets $C$ for $\lambda$ as
\begin{equation}
\Pi (C)=\int_{\sigma(A)\cap C} dE(\lambda).
\label{112}
\end{equation}

All this can be seen as the general epistemic interpretation of quantum states and projection operators. It is related to the QBist interpretation, but is more general. It can also be seen as a concrete specification of the Relational Quantum Mechanics and of interpretations related to information.

In (\ref{111}, \ref{112}) $\lambda$ may be seen as a maximally accessible variable. If $\theta$ is another maximally accessible variable, it will be associated with another operator $B$, and $A$ and $B$ will not be commuting. We can then say that $\lambda$ and $\theta$ are complementary variables in the sense of Bohr. In a physical context, Niels Bohr's complementarity concept has been thoroughly discussed by Plotnitsky [42]. 

Here is Plotnitsky’s definition of complementarity:

(a)	a mutual exclusivity of certain phenomena, entities, or conceptions; and yet

(b)	 the possibility of applying each one of them separately at any given point; and

(c)	the necessity of using all of them at different moments for a comprehensive account of the totality of phenomena that we consider.

This definition points at the physical situation discussed above, and has Niels Bohr’s interpretation of quantum mechanics as a point of departure. However, in my opinion the definition can also be carried over to a long range of macroscopic phenomena or conceptions. In particular, the concept is useful in connection to quantum cognitive modeling [28, 29, 43] and in quantum decision theory, see Yukalov and Sornette [44, 45, 46, 47, 48].

\section{Focused likelihood}

Assume that we have focused on some index $a\in\mathcal{A}$ and on an experiment with parameter $\theta^a$ and data $x^a$. In much of the quantum theory literature, $\theta^a$ is discrete, and one does not distinguish between $\theta^a$ and $x^a$. However I will think of $\theta^a$ as a general parameter, typically continuous, and I will model the uncertainty of the measurement apparatus by a statistical model. Then this statistical model gives a likelihood $f(x^a |\theta^a )$, and we can define the likelihood effect
\begin{equation}
F^a(x^a )=\int_{\sigma(A^a)} f(x^a |\theta^a =u )dE^a (u) ,
\label{likeff}
\end{equation}
where $\sigma(A^a)$ is the spectrum of the operator $A^a$ corresponding to $\theta^a$, and $E^a$ is the corresponding projection valued measure. I rely here on the discussion of Section 4.4. This is a generalization of an ordinary definition in quantum theory, where any operator $F=\sum_j p_j \Pi_j$ so that $0\le p_j \le 1$ and $I=\sum_j \Pi_j$ is a resolution of the identity, is called an effect. Note that the discrete version of (\ref{likeff}) is $F^a(x^a)=\sum_j p(x^a|\theta^a=u_j^a)\Pi^a (u_j^a )$, where $p(x^a|\theta^a =u_j^a)$ is a point probability, so this is obviously an effect. That the general likelihood is an effect, follows by a limiting argument.

The likelihood principle of statistical inference (see Subsection 10.1) says all experimental information from a given experiment $a$ is contained in the likelihood $f(x^a|\theta^a)$. In [1] a discrete version of a focused likelihood principle was proved under a certain regularity condition. Here I formulate the more general principle:
\bigskip

\textbf{The focused likelihood principle.} \textit{Consider the choice of maximally accessible experiments $a$ that are subject to the same inaccessible variable $\phi$. Let $q^a (\theta^a|x^a)$ be the experimental evidence from experiment $a$, for instance a posterior probability. Then, taking into account the choice of $a$, $q^a$ is a function of the likelihood effect: $q^a=q(F^a(x^a))$.}
\bigskip

\textbf{Theorem 6.} \textit{The focused likelihood principle follows from the ordinary likelihood principle.}
\bigskip

\textit{Proof.} I will prove the discrete version; the general case follows by an extension of the argument. So let $F^a(x^a)=\sum_j p(x^a|\theta^a=u_j^a)\Pi^a (u_j^a )$, where $\Pi^a (u_j^a)$ is the projection upon the eigenspace of the operator $A^a$ corresponding to the eigenvalue $u_j^a$. In many applications $u_j^a$ can be chosen to be independent of $a$. If not, let $\{u_j\}$ be the union of all $u_j^a ; a\in\mathcal{A}$, and define $\Pi^a (u_j)=0$ if $u_j$ is not an eigenvalue of $A^a$. Then we can write $F^a(x^a)=\sum_j p(x^a|\theta^a=u_j)\Pi^a (u_j )$. Now assume that $F^a(x^a)=F^b(x^b)$ for two indices $a$ and $b$. Then it follows from Proposition 5.5 in [1] that the eigenspaces can be ordered such that 1) $\Pi^a(u_j)=\Pi^b(u_j); j=1,2,...$; 2); 2) $p(x^a|\theta^a=u_j)=p(x^b|\theta^b=u_j)$ when both are defined. From 1) it follows that the experimental questions connected to $a$ and $b$ are coinciding. 2) together with the ordinary likelihood principle implies that the experimental evidences are equal. $\Box$

 \section{The Born formula}
 
 I start with the discrete case and a maximally accessible variable $\theta^a$ defined in connection to an agent $B$ and a physical system $S$. It is crucial that $\theta^a$ is maximally accessible, and that it has already been accurately measured by $B$ to a value $\theta^a = u_i^a$. By the theory already considered here, this gives a unique state $|a;i\rangle$.
 
 Now consider another variable $\theta^b$ connected to $B$ and $S$. Born's formula, the basis for quantum probability calculations, gives the probability distribution of $\theta^b$ in this situation.
 
 As in [1] I make some assumptions coupled to the agent $B$. I do not see $B$ as being perfectly rational, but during his experiment he adhers to certain ideals, and these ideals can be modeled by a superior actor $D$ which is perfectly rational. I assume that $D$ has priors found in some way, so that he can do a Bayesian analysis. The experimental evidence connected to $D$ is then given by his posterior probabilities $q=q^a$ depending on the data $x^a$. The focused likelihood principle, specialized to a single experiment, states that the experimental evidence, including the questions related to this evidence, must be a function of the likelihood effect. Thus $q^a$ must be a function of the likelihood effect $q^a=q(F^a)$.
 
 Note that I do not assume that $B$ has a prior and is able to do a Bayesian analysis, only that the ideal actor $D$ is able to act as a Bayesian.
 
 The perfect rationality of $D$ is formulated in terms of hypothetical betting situations and
 \bigskip
 
 \textbf{The Dutch Book Principle.} \textit{No choice of payoffs in a series of bets shall lead to a sure loss for the bettor.}
 \bigskip
 
 The setting described above is called a rational epistemic setting. It is assumed that the (focused) likelihood principle holds. The following theorem is proved in [1]:
 \bigskip
 
 \textbf{Theorem 7.} \textit{Assume a rational epistemic setting. Let $F_1 , F_2 ,...$ be likelihood effects in this setting, and assume that $F_1+F_2 +...$ also is a likelihood effect. Then}
 \begin{equation}
 q(F_1 +F_2 +...)=q(F_1)+q(F_2)+....
 \label{q}
 \end{equation}
 
 When $F_1, F_2,...$ belong to the same experiment, this is quite obvious, but it is proved also for the case where the experiments are different. A function $q(\cdot)$ satisfying the countable additivity condition (\ref{q}) and in addition $0\le q(F)\le 1$ for all $F$ and $q(I)=1$, is called a generalized probability measure. In [49, 50] it is called a frame function. I can now make use of the following theorem by Busch [51], cp. also [49]:
 \bigskip
 
 \textbf{Theorem 8.} \textit{Any generalized probability measure is of the form $q(F)=\mathrm{trace}(\sigma F)$ for some density operator $\sigma$.}
 \bigskip
 
 The arguments above are valid for any sets of experiments in a rational epistemic setting, where $q=q(F)$ is some probability, a function of the likelihood effect, which in some way expresses the experimental evidence of the experiment connected to the likelihood effect $F$.
 
Define now a perfect experiment as one where the measurement uncertainty can be disregarded. The quantum mechanical literature operates very much with perfect experiments which give well-defined states $|j\rangle$. From the point of view of statistics, if, say the 99\% confidence or credibility region of  $\theta$ is the single point $u_j$, we can infer approximately that a perfect experiment has given the result $\theta=u_j$.

Assume now that we have prepared the system by a perfect experiment giving the result $\theta^a =u_i$ for some accessible variable $\theta^a$. Let us then do a new perfect experiment, asking the question `What is the value of $\theta^b$?' for a new accessible variable $\theta^b$. Assume to begin with that both the accessible variables are maximal, so that the relevant eigenvalues are non-degenerate with eigenvectors $|a;i\rangle$, respectively $|b;j\rangle$. Then the following is proved in [1] from Theorem 8:
\bigskip

\textbf{Theorem 9. Born's formula.} \textit{Assume a rational epistemic setting. In the above sitiuation we have}
\begin{equation}
P(\theta^b =u_j^b |\theta^a =u_i^a ) = |\langle a;i |b;j\rangle |^2 .
\label{Born1}
\end{equation}
\smallskip

There are several generalizations of this Born formula. One can let the last eigenvalue be degenerate with projector $\Pi_j^b$ on the eigenvector space corresponding to $\theta^b =u_j^b$. Then the righthand side of (\ref{Born}) should be replaced by $\langle a;i| \Pi_j^b |a;i\rangle$. And the first preperatorial experiment can be one where we only have probabilities for the different values $u_i^a$, resulting in a density operator $\sigma^a$. Then $P(\theta^b=u_j^b|\sigma^a ) =\mathrm{trace}(\Pi_j^b \sigma^a)$.

It is of some interest that Theorem 8, which is the basis, also is valid for non-perfect experiments. Following the same arguments, this leads to (\ref{p}) and (\ref{M}) below, which can be related to quantum measurement theory. 

Finally, the Born formula can be extended to the continuous case. In order to formulate the most general case, let the system be prepared in a mixed state with density $\sigma^a = \int \pi(\theta^a )dE^a (\theta^a )$, and let us consider an experiment with a new accessible variable $\theta^b$ whose operator has a spectral distribution $dE^b (\theta^b )$. The data $x$ of the experiment has a statistical model which gives $P(x\in C|\theta^b)$ for sets $C$ in the sample space. It is assumed that the model is dominated: $P(x\in C|\theta^b) = \int_C f(x|\theta^b )d\xi (z)$ for a likelihood $f(x|\theta)$ and a measure $\xi$ on the sample space. Then one can define the likelihood effect
\begin{equation}
F^b (x)=\int_{\theta^b} f(x|\theta^b)dE^b (\theta^b ),
\label{F}
\end{equation}
and the focused likelihood principle holds. The continuous Born formula may be written
\begin{equation}
f(x|\sigma^a )= \mathrm{trace} (F^b (x)\sigma^a ).
\label{cborn}
\end{equation}

Note that $\int F^b (x) d\xi (x) =I$, so that one can define an positive operator valued measure $M$ by
\begin{equation}
M(C)=\int_C F^b (x) d\xi (x) =\int_{\theta^b} P(X\in C|\theta^b )dE^b (\theta^b ).
\label{MM}
\end{equation}

As an application of the discrete version of Born's formula, we give the transition probabilities for spin 1/2 particles. I will, for a given direction $a$ define $\theta^a =+1$ if the measured spin component by a perfect measurement for some given particle is $+\hbar /2$ in this direction, $\theta^a =-1$ if the component is $-\hbar /2$. Assume that $a$ and $b$ are two directions in which the spin component can be measured.
\bigskip

\textbf{Proposition 1.} \textit{For a spin 1/2 particle we have}
\begin{equation}
P(\theta^b =\pm 1 |\theta^a =+1)=\frac{1}{2} (1\pm \mathrm{cos}(a\cdot b)).
\label{spincomp}
\end{equation}

This is proved in many textbooks, and also in [30].
\bigskip

It is interesting that a similar derivation of the Born rule from frame functions is discussed in [49, 50]. In [50] a more abstract setting is considered, so that in addition to the Born rule, the corresponding state-updating rule (the collapse rule) can be derived from the same setting. It is stated in the introduction to [50] that the authors want to breathe new life into the knowledge-based view of quantum theory. In the present paper I share the same ambition; in my vocabulary, knowledge-based is the same as epistemic.

The derivations of Sections 4-9 now give a starting point for the development of quantum theory as it is given in many textbooks and also in Chapter 5 of [1]. The Schr\"{o}dinger equation is also developed from an epistemic point of view in [1].

\section{Some statistical inference theory}

Statistical inference theory was mentioned on several occations above. I now give an introduction to 3 approaches to statistical inference. After this, I will discuss an example and some brief elements of quantum measurement theory, in order to illustrate more closely the connections between quantum theory and statistical inference theory, as I see it.

\subsection{Frequentist and Bayesian inference}

The basic notions of statistical inference are data $X$, varying in a space $\Omega_X$, a vector or scalar parameter $\theta$ varying in a space $\Omega_\Theta$ and a statistical model defined as follows: There is a $\sigma$-algebra $\mathcal{E}_X$ of subsets of $\Omega_X$, and for each $\theta\in\Omega_\Theta$ there is a  probability measure $P_X^\theta$ on the measurable space $(\Omega_X ,\mathcal{E}_X )$. Usually one assumes that this family of probability models is dominated: There exist a measure $\xi$ on $(\Omega_X ,\mathcal{E}_X )$ and a likelihood function $f(x|\theta)$ such that $dP_X^\theta =f(x|\theta)d\xi$.

In statistics, $\theta$ is thought to model the unknown feature that one is interested in, the state of the system in question. The data $X$ are the potential observations, and the purpose of the modeling is multifold: To give a rough description of the data generating process; provide parameters that can be estimated from data; allow focusing on certain parameters; give a language for asking questions about nature, and give a possibility to study deviations from the model and choosing new models.

Here I will concentrate on asking questions about nature, in particular on estimating $\theta$ from observations. Given observed data $X=x$, one makes a decision $\delta(x)$, where the function $\delta$ is thought to vary on some action space $\Omega_A$. From this point of view, one can say that statistical inference is the science of finding an optimal $\delta$ in concrete problems.

The simplest concept is that of point estimation: The parameter $\theta$ is estimated by a function of the data: $\hat{\theta}(x)$. The properties of this estimation procedure is evaluated  by looking at the before-experimental situation and using the statistical model: With the stochastic variable $X$ inserted, $\hat{\theta}(X)$ is called an estimator. One good property might be that the estimator is unbiased: $E^\theta (\hat{\theta}(X))=\theta$ for all $\theta$, exactly or perhaps approximately. Another good property may be that it has a small variance. These two properties are sometimes combined in the requirement that the estimator should have a mean square error that is as small as possible, where the mean square error is defined as
\begin{equation}
MSE(\hat{\theta}(X))=E^\theta( (\hat{\theta}(X)-\theta )^2)=\mathrm{Var}^\theta (\hat{\theta}(X))+(E^\theta(\hat{\theta}(X)-\theta))^2.
\label{MSE}
\end{equation}

In a more general decision situation one may have defined a loss function $\gamma (\theta,\delta(x))$ that one may wish to minimize in some sense. The first step is then to define the risk
\begin{equation}
R(\delta, \theta)=E^\theta (\gamma(\theta, \delta(X))).
\label{risk}
\end{equation}
A basic problem is that the risk in almost all cases cannot be minimized uniformly for all $\theta$. Which way one proceeds with this problem, depends on what school one wants to adhere to. One way to go on, is to limit oneself in the point estimation case to unbiased estimators; for this theory and related theories, see [52].

A related problem on the case of a real-valued $\theta$ is to find an interval from data in which one believes that $\theta$ belongs. Again the solution depends on what school one wants to adhere to. I first describe the solution of the frequentist school.

One is in a situation where the data are given, but still one wants to use the statistical model as if one was in a pre-experimental situation. This presupposes an hypothetical future experiment completely equivalent to the experiment that one has performed, and that one evaluates probabilities connected to this future experiment. Given a confidence coefficient $1-\alpha$, say 0.95, the aim is then to find lower and upper estimators $\underline{\theta}$ and $\bar{\theta}$ such that
\begin{equation}
P^\theta(\underline{\theta}(X)\le\theta\le\bar{\theta}(X))=1-\alpha.
\label{confidence}
\end{equation}
One then reports a confidence interval $[\underline{\theta}(x),\bar{\theta}(x)]$.

Again a related problem is that of hypothesis testing, say that one wants to test a null hypothesis $\theta=\theta_0$. An important concept is then that of the alternative hypothesis. Say that one considers a one-sided alternative $\theta > \theta_0$, has found a point estimator $\hat{\theta}$ for $\theta$ and the value $\hat{\theta}(x)$ for the experimental data at hand.  One may then report a $p$-value
\begin{equation}
 p=P^{\theta_0}(\hat{\theta}(X)>\hat{\theta}(x)), 
\label{pvalue}
\end{equation}
refering to a hypothetical experiment under the null hypothesis, giving the result $X$. If the p-value is small, a traditional limit has been 0.05, the finding of the experiment is reported as being significant.

The p-values have been very much used in empirical research, and this use has been heavily attacked recently. One problem is that when many questions are addressed, in one or in several studies, the overall probability of getting a `significant' result is large. The conclusion is that one should be very careful with the automatic use of p-values. Often the report of a confidence interval will be more informative.

In concrete experiments one often has data $X$ that consists of several independent observations, in general one may have have very much data. One mechanism by which one can reduce data, is related to the concept of sufficiency. A function $T$ of the data is called a statistic. A statistic $T$ is called sufficient if the distribution of $X$, given $T$ and $\theta$, is independent of $\theta$. Under weak conditions one then has factorization of the likelihood
\begin{equation}
f(x|\theta)=h(x)g(T(x)|\theta).
\label{factorization}
\end{equation}
Thus the essential part of the observation is T(x); apart from this, the data from the model could have been obtained by independent computer simulation. This is related to the sufficiency principle: If data $x_1$ and $x_2$ have $T(x_1)=T(x_2)$, then $x_1$ and $x_2$ contain the same experimental evidence about $\theta$. Here the concept `experimental evidence' is left undefined.

Birnbaum showed (see [1]) that the sufficiency principle together with an equally intuitive conditionality principle imply the likelihood principle: If $x_1$ and $x_2$ have proportional likelihood functions, where the constant of proportionality is independent of $\theta$, then $x_1$ and $x_2$ contain the same experimental evidence about $\theta$. Thus all information about $\theta$ is contained in the likelihood function.

There have been discussions also around the likelihood principle. In [1] the view is advocated that the likelihood principle should be taken as conditional, given the context of the experiment.

A completely different approach to statistical inference is given by the Bayesian school. Bayesian inference is possible if one has a prior distribution $P(d\theta )$ of the parameter. Then one can define a Bayesian risk
\begin{equation}
B(\delta )=\int R(\delta ,\theta ) P(d\theta),
\label{Bayesrisk}
\end{equation}
where $R(\delta,\theta )$ is given by (\ref{risk}), and the problem is simply to find the decision rule that minimizes $B(\delta)$. In the case of point estimation and quadratic loss function for given data, this leads to
\begin{equation}
\hat{\theta}(x)=\int \theta P(d\theta |x),
\label{Bayestheta}
\end{equation}
where
\begin{equation}
P(d\theta |x)=\frac{ l(x|\theta)P(d\theta )}{\int l(x[\eta)P(d\eta )}.
\label{posterior}
\end{equation}
The distribution (\ref{posterior}) is called the posterior distribution of $\theta$, and is a fundamental tool of Bayesian inference. (\ref{posterior}) is an instance of Bayes' formula.

Under the posterior distribution, the parameter $\Theta$ is a random quantity, and (\ref{Bayestheta}) may be written $\hat{\theta}(x)=E(\Theta |x)$. 

The Bayesian concept corresponding to a confidence interval is that of a credibility interval: Assume that one can find $\theta_* (x)$ and $\theta^*(x)$ such that
\begin{equation}
P(\theta_* (x)\le \Theta \le\theta^*(x)|X=x)=1-\alpha.
\label{credibility}
\end{equation}
Then $[\theta_*(x),\theta^*(x)]$ is called a credibility interval for $\Theta$ with credibility coefficient $1-\alpha$. Note that, in contrast to (\ref{confidence}), the probabilities are now computed directly from the posterior distribution of $\Theta$.

The main problem of the Bayesian approach to inference is to find a prior $P(d\theta )$. A subjective Bayesian will say that this always can be specified by the user; sometimes one here makes use of a hypothetical betting construction.
Another school relies on an `objective' prior; one such approach assumes that there is a group defined on the parameter space. 

Sometimes, but not always, one starts with a group $G^*$ acting on the sample space $\Omega_X$, so that this induces a group $G$ on the parameter space by assuming that the model parameter is uniquely determined by the model: $g\theta$ is defined by $P^{g\theta}_X=P^\theta_{gX}$. This defines a group homomorphism, sometimes an isomorphism. Also, there is a group $G^{**}$ acting on the action space $\Omega_A$, and it is assumed that there is an homomorphism from $G^*$ to $G^{**}$ defined through an permissible inference rule: $g^{**}\delta(x)=\delta(g^*x)$. If the action is an estimation procedure, and the homomorphism from $G^*$ to $G$ is an isomorphism, it is natural to assume also that this last homomorphism is an isomorphism. Also, one usually assumes that the loss function satisfies $\gamma(g\theta, g^{**}a)=\gamma(\theta, a)$. In the isomorphism case, I will use the symbols $G,g$ instead of $G^*,g^*$ and $G^{**}, g^{**}$. 

An estimator $\hat{\theta}(X)$ is called equivariant if it transforms under the group in the same way as the parameter $\theta$, i.e., $g\hat{\theta}(X)=\hat{\theta}(g^* X)$. More generally, a decision rule is called invariant if $g^{**}\delta(x)=\delta(g^* x)$ for all $x$. The risk function (\ref{risk}) of an invariant decision rule is constant on orbits of $\theta$, see [58]. In the transitive case this leads to a unique invariant decision rule, see also [53]. 

When a group $G$ is defined on the parameter space, there are many arguments to the effect that the right invariant measure connected to $G$ should be used as an objective prior for $\theta$. (See [30], for instance.) An argument sometimes raised against this, is that it very often leads to improper priors. As recently shown in [53], however, this problem can be solved by a slight relaxation of Kolmogorov's axioms if the marginal law of $X$ is $\sigma$-finite: There are events $B_1 , B_2 ,...$ with $\Omega_X =\cup_i B_i$ and $P(X\in B_i)<\infty$ for $i=1,2,...$.

For comparing statistical inference theory with quantum theory, we need to consider the case where both the parameter and the data are discrete. Then, with enough data, the credibility interval or the confidence interval may shrink to a single point $\theta_0$, so that $P(\Theta =\theta_0 |X=x)=1-\alpha$, respectively $P(\hat{\theta}(X)=\theta_0)=1-\alpha$. When $\alpha$ is small enough, we may then identify the post-experimental parameter value with $\theta_0$, and thus disregard experimental error. This explains why there is no distinction between data and parameters in quantum mechanical textbooks. 

A classical statistical text treating both Bayesian and frequentist inference, is Berger [35]. Otherwise, there are many textbooks on statistical inference, for instance Bickel and Doksum [54]. Ferguson [55] is a classical text using a decision theoretical approach. For more on the case where a group is defined on the parameter space, see [30] and [56]. One can also refer to the discussion in [1].

\subsection {Fiducial theory}

For an introduction to fiducial inference, the third approach to statistical inference, see for instance [57]. Fiducial inference as an alternative to Bayesian inference was first proposed by Fisher [58], has been in discredit for many years, but has been the subject to several investigations recently. I use some space to it here because it is a distinctly new approach to statistical inference, different from frequentist and Bayesian inference. And the use in [57] of a transitive group defined on the parameter space is similar to one of my assumptions behind my approach to quantum mechanics.

The connections between optimal statistical inference and fiducial models are also studied in [57], and this is illustrated by several examples.

\subsection{Connection between approaches to statistical inference}

For an outsider it may be confusing that there are several schools of statistical inference. Then it is reassuring that there are connections between these schools. First, in cases where there is very much data, the different theories give essentially the same result. In these cases, the effect of the choice if prior vanishes almost completely.

An even closer connection between the theories exists when there is a transitive group $G$ defined on the parameter space. Then there is a classical result by Fraser [59, 60] and generalized by Taraldsen and Lindqvist [61] saying that the fiducial distribution coincides with the Bayesian posterior from the right invariant measure as a prior.

There is also a close connection between Bayesian and frequentist inference in this case: In [30] the following general result is proved (recall the definition of a proper group in Section 2. An estimator is equivariant if it transforms under $G$ in the same way as the corresponding parameter): 
\bigskip

\textbf{Proposition 2.} \textit{Assume that $G$ is proper and is acting transitively both on $\Omega_X$ and on $\Omega_\Theta$. Consider frequentist inference on one hand and Bayesian inference with right invariant prior on the other hand. Let $\eta(\theta)$ be a one-dimensional continuous parametric function, and let $\hat{\eta}_1(x)$ and $\hat{\eta}_2 (x)$ be two equivariant estimators of $\eta(\theta)$ such that $\hat{\eta}_1(x)<\hat{\eta}_2(x)$ for all $x$. Define $C(x)=\{\theta: \hat{\eta}_1(x)\le\theta\le \hat{\eta}_2(x)\}$. Then $C(x)$ is a credibility interval and a confidence interval of $\theta$ with the same credibility coefficient/ confidence coefficient.}
\bigskip

These results seem to give a special status to `objective' Bayesian inference based on right invariant prior.

 \section{A macroscopic example}
 
 Going back to quantum theory, a very relevant question is now: Are all the results of Sections 4-9, including Born's formula, by necessity confined to the microworld? Recently, physicists have also become 
interested in larger systems where quantum mechanics is valid, see [62]. Of even more interest are the quantum models of cognition, see [28,29], quantum models in finance (e.g. [43, 63]) and Quantum Decision Theory [44-48].  As I have defined it, there is nothing microscopic about the epistemic setting. It may or may not be that the assumptions made above  also are valid for some larger scale systems. The following example illustrates the point.
\bigskip

In a medical experiment, let $\mu_{a}, \mu_{b}, \mu_{c}$ and $\mu_{d}$ be continuous inaccessible parameters, the hypothetical effects of treatment $a, b, c$ and $d$, 
respectively. Assume that the focus of the experiment is to compare treatment $b$ with the mean effect of the other treatments, which is supposed to give the parameter 
$\frac{1}{3}(\mu_{a}+\mu_{c}+\mu_{d})$. One wants to do a pairwise experiment, but it turns out that the maximal parameter which can be estimated, is
\[\theta^b =\mathrm{sign}(\mu^b -\frac{1}{3}(\mu_{a}+\mu_{c}+\mu_{d})).\]
(Imagine for example that one has four different ointments against rash. A patient is treated with ointment $b$ on one side of his back; a mixture of the other 
ointments on the other side of his back. It is only possible to observe which side improves best, but this observation is assumed to be very accurate. One can in 
principle do the experiment on several patients, and select out the patients where the difference is clear.)
This experiment is done on a selected set of experimental units, on whom it is known from earlier accurate experiments that the corresponding parameter
\[\theta^a =\mathrm{sign}(\mu^a -\frac{1}{3}(\mu_{b}+\mu_{c}+\mu_{d}))\]
takes the value $+1$. In other words, one is interested in the probabilities
\[\pi =P(\theta^b=+1|\theta^a=+1).\]

Consider first a Bayesian approach. Natural priors for $\mu_{a},...,\mu_{d}$ are independent normal distributions, $N(\nu, \sigma^2)$ with the same $\nu$ and $\sigma$. By location 
and scale invariance, there is no loss in generality by assuming $\nu =0$ and $\sigma=1$. Then the joint prior of 
$\zeta^a =\mu_{a}-\frac{1}{3}(\mu_{b}+\mu_{c}+\mu_{d})$ and $\zeta^b =\mu_{b}-\frac{1}{3}(\mu_{a}+\mu_{c}+\mu_{d})$ is multinormal with mean 
$\bm{0}$ and covariance matrix
\[\left( \begin{array}{cc}\frac{4}{3}&-\frac{4}{9}\\
-\frac{4}{9}&\frac{4}{3}\end{array}\right).\]
(A vector variable $\bm{v}$ has a multinormal distribution with mean $\bm{\mu}$ and covariance matrix $\bm{\Sigma}$ if $\bm{a}'\bm{v}$ is $N(\bm{a}'\bm{\mu}, \bm{a}'\bm{\Sigma}\bm{a})$ for every constant vector $\bm{a}$.)
A numerical calculation from this gives
\[\pi=P(\zeta^b>0|\zeta^a>0)\approx 0.43.\]
This result can also be assumed to be valid when $\sigma\rightarrow\infty$, a case which in some sense can be considered as independent objective priors for 
$\mu_{a},...,\mu_{d}$.

Now consider a quantum theory approach to this experiment.  Since again scale is irrelevant, a natural group on $\mu_{a},...,\mu_{d}$ is a 4-dimensional rotation group 
around a point $(\nu,...,\nu)$ together with a translation of $\nu$. Furthermore, $\zeta^a$ and $\zeta^b$ are contrasts, that is, linear combinations with coefficients 
adding to 0. The space of such contrasts is a 3-dimensional subspace of the original 4-dimensional space, and by a single orthogonal transformation, the relevant subset 
of the 4-dimensional rotations can be transformed into the group $G$ of 3-dimensional rotations on this latter space, and the translation in $\nu$ is irrelevant. One such 
orthogonal transformation is given by
\[\psi_{0}=\frac{1}{2}(\mu_{a}+\mu_{b}+\mu_{c}+\mu_{d}),\]
\[\psi_{1}=\frac{1}{2}(-\mu_{a}-\mu_{b}+\mu_{c}+\mu_{d}),\]
\[\psi_{2}=\frac{1}{2}(-\mu_{a}+\mu_{b}-\mu_{c}+\mu_{d}),\]
\[\psi_{3}=\frac{1}{2}(-\mu_{a}+\mu_{b}+\mu_{c}-\mu_{d}).\]
Let $G$ be the group of rotations orthogonal to $\psi_{0}$. We find
\[\zeta^a=-\frac{2}{3}(\psi_{1}+\psi_{2}+\psi_{3}),\]
\[\zeta^b=-\frac{2}{3}(\psi_{1}-\psi_{2}-\psi_{3}).\]
The rotation group element transforming $\zeta^a$ into $\zeta^b$ corresponds under $G$ to the rotation group element $g_{ab}$ transforming 
$a=-\frac{1}{\sqrt{3}}(1,1,1)$ into $b=-\frac{1}{\sqrt{3}}(1,-1,-1)$ in a simpler rotation group, which may be taken as a rotation of the Bloch sphere for a spin 1/2 system. In conclusion, the whole situation is completely equivalent to the spin-example of 
Proposition 1 in Section 9 and satisfies the assumptions of the symmetrical epistemic setting. Making the rationality assumption  then implies:
\[\pi=P(\mathrm{sign}(\zeta^b)=+1|\mathrm{sign(\zeta^a)=+1})=\frac{1}{2}(1+a\cdot b)=\frac{1}{3}.\]
\smallskip

I guess that many statisticians will prefer the Bayesian calculations here for the quantum theory calculations. But the prior chosen in this example must be considered somewhat arbitrary, and its 'objective' limit may lead to conceptual difficulties. 

The purpose of this example is not primarily to show how quantum theory applies in a macroscopic setting. Other, perhaps more interesting examples can be found in cognitive modeling  [28, 29] and in economics (e.g., [63]). The purpose is more to indicate a border area where quantum modeling may or may not be valid, i.e. where the assumptions made in this article may or may not hold. And to show in a precise example how one aspect of the different ways of looking at uncertainty in statistical modelling may in principle be illustrated by a quantum theory calculation.

 \section{Some quantum measurement theory}
 
 Consider first a simple measurement of a maximally accessible discrete variable $\theta$ with associated operator $A=\sum_n u_n |n\rangle\langle n|$. Let the statistical model for the measurement be given by $P(X=x_j|\theta=u_n )$, assuming discrete data. Then assuming that the physical system first is prepared in some  mixed state $\sigma$, one makes a measurement on $\theta$, and the probability of obtaining the measurement result $X=x_j$ is according to the data version of Born's formula
 \begin{equation}
 p_j =P(X=x_j|\sigma)=\mathrm{trace} (M(j)\sigma ),
 \label{p}
 \end{equation}
 where
 \begin{equation}
 M(j)=\sum_n P(X=x_j|\theta=u_n )|n\rangle\langle n|.
 \label{M}
 \end{equation}
Then the final state must in some way be determined by $M(j), \sigma$ and $p_j$. Note that trivially $\sum_j M(j) =I$. The formula for the final state can be made consistent with the measurement theory of [64], summarized below, if one can find operators $A_j$ such that $M(j)=A_j^{\dagger}A_j$.
 
 This discontinuous change of state has caused much discussion in the literature, but with an epistemic interpretation of quantum theory it seems to be less problematic: It just represents an updating of the information that one one has on $\theta$; deeper reasons for such updating are given in [50].
 
 According to [64], the most general measurement one can have on a system can be described as follows: Let $\{A_j\}$ be a set of operators satisfying $\sum_j A_j^{\dagger} A_j=I$. We want to measure a variable $\theta$ with possible values $u_n$. If $\sigma$ is the state of the system before measurements, $\tilde{\sigma}_j$ is the state of the system after obtaining the measurement result $z=z_j$, and $p_j$ is the probability of obtaining this result, then 
 \begin{equation}
 \tilde{\sigma}_j =\frac{A_j \sigma A_j^{\dagger}}{p_j},
 \label{sigma1}
 \end{equation}
 \begin{equation}
 p_j =\mathrm{trace}(A_j^{\dagger}A_j \sigma ).
 \label{p1}
 \end{equation} 
 
 In general, (\ref{sigma1}) and (\ref{p1}) give the inference rule implied by the quantum measurement. The arguments behind these formulae, as given in [64], is of interest. I will sketch a possible starting point of these arguments from a more statistical perspective.
 
 From this perspective, the measurement is described by the data $X$, and $\theta$ is the parameter of interest. During measurement one goes from the unknown variable $(X,\theta)$ to the known variable $(X=x, \hat{\theta}(x))$. This transition can be described by a group element $h\in G\otimes G$, and $h$ can be represented by a unitary operator $U=U(h)$ acting on the vectors $|X=x_i\rangle |\theta=u_j\rangle$. In [64] similar arguments are given, using the concepts of probe and target. The existence of the unitary operator $U$ can also be motivated by a classical theorem by Wigner [65].
 
 In this language the derivation in [64] starts by conditioning on the data  $X=x$, giving as a combined initial state  the density matrix $\sigma_{comb}=|0\rangle\langle 0|\otimes \sigma$. The operator $U$ is then written in terms of sub-blocks acting on the parameter space as $U=\sum_{jj'}[j\rangle\langle j'|\otimes A_{jj'}$, and it is argued that the first column of these blocks, $A_j =A_{j0}$ satisfy the resolution of the identity $I=\sum_j A_j^{\dagger}A_j$, and from this and from the initial state the final state (\ref{sigma1}) is derived. Alternatively, one could consider the derivation of the corresponding formula in [50].
 
 The special case when $\sigma=\sum P(n)|n\rangle\langle n|$ and all the operators $A_j$ are diagonal in the vectors $|n\rangle$, i.e., $A_j =\sum_n A(j,n)|n\rangle\langle n|$, is also discussed in [64]. This case is shown to be equivalent with Bayesian inference with priors for $\theta$ equal to $P(n)$ and likelihood function $P(j|n)=|A (j,n)|^2$.
 
 The point of this brief recapitulation of quantum measurement theory is to indicate that quantum measurement is a generalization of Bayesian inference, and that the prior in this Bayesian inference theory could be anything, subjective or objective. The Bayesian inference corresponds to the case where everything can be formulated in terms of the projectors $|n\rangle\langle n|$, i.e., in a universe where we are only interested in a single resolution of the identity $I=\sum_n |n\rangle\langle n|$, thus a single question posed to nature. In particular, then $M(j)=A_j^{\dagger}A_j$ is diagonal in the $\{|n\rangle\}$ basis, as indicated in (\ref{M}). 
 
 There are many further aspects of quantum measurement theory, aspects that are not discussed in this paper.

 \section{Discussion}
 
 The developments given in Sections 4-9 here represent an alternative approach towards quantum theory, and these discussions should replace parts of Chapter 4 in my book [1]. The discussions are rather technical, but the foundation introduced based on conceptual inaccessible and accessible conceptual variables, seems to be much more intuitive than starting directly with the ordinary formal Hilbert space foundation.
 
 The resolutions of the identity (\ref{6},\ref{9}, etc.) are crucial to this approach. They can be said to represent questions to nature: `What is the value of $\theta$?' for an associated maximally accessible variable $\theta$ , and at the same time they give positive-operator valued measures. The resolution (\ref{6}) is special, since the operators involved are not orthogonal projections.\footnote{By Neumark's dilation theorem, the POVM here can be `lifted' to a projection-valued measure.} In most other cases discussed here, the resolutions of the identity give directly projection-valued measures. Selecting one concrete projector $\Pi$ (eigenspace $V$ for a corresponding operator) gives a sharp answer to the relevant question to nature.The probabilities associated with these answers are given by the Born rule $p=\mathrm{trace}(\Pi \sigma)$ if we start with a state defined by the density operator $\sigma$. Different resolutions of the identity correspond to complementary questions.
 
 The way I use `question' here, a question may be composed of several elementary questions. This corresponds to the case where $\theta$ is a vector, and the operators corresponding to the elementary questions commute.
 
 As mentioned in the previous Section, ordinary Bayesian inference in a quantum measurement situation corresponds to the case where we only have a single resolution of the identity. This is true in the discrete case, but it can be generalized. As thoroughly discussed in Section 10, there are also other approaches to statistical inference than the Bayesian one. It is interesting that both frequentist inference and fiducial inference are connected to `objective' Bayesian inference, the one assuming a transitive group acting on the parameter space and the use of a right invariant prior. 
 
 Group theory, the assumption of a transitive group acting on the variable (parameter) space, is also important for the approach towards quantum theory advocated here. If we have a group defined on this space which is not transitive, model reduction - reduction to an orbit of the group - is called for. This is the approach to quantization advocated here. It is interesting that a similar model reduction also is useful in ordinary statistical inference.
 
 The example in Section 11 shows that a quantum mechanical approach and a Bayesian statistical approach in general may give different solutions to an inference problem. 
 
 In [28, 29] it is proposed to use quantum theory in cognitive modeling and in decision theory. This use of quantum theory is important, but it is not discussed directly in the present paper. A parallel discussion may be provided, however. Conceptual variables are used in making decisions. These variables may be inaccessible if they are so comprehensive that no clear decision can be made. Then it may be a solution to focus on simpler, accessible decision variables.
 
 The relationship between my approach and QBism (see for instance [23, 26, 66] on this) was briefly discussed in Secton 2. I also regard the Bayesian way of thinking as useful, in particular when we talk about an ideal observer. But I look upon Bayesianism as much wider than subjective Bayesianism; as mentioned above, `objective' Bayesianism based on a right invariant prior is useful, both in applications and when discussing connections between different approaches to inference.
 
 A QBist advocates subjective Bayesianism and considers hypothetical betting constructions to be a basis for making decisions. I rather regard `decision' as a more primitive concept. A decision can be based on desires, beliefs or knowledges in some combinations, but we very often make decisions without having neither the time nor the opportunity to think of a hypothetical betting situation. See also the discussion of Quantum Decision Theory in [44-48].
 
 When asking questions to nature, two kinds of decisions are important. First we must decide how to focus before asking the question, and next, after an answer is obtained, we must decide how to interpret the answer.
 
 \section{The epistemic interpretation}

Consider a physical system, and an observator  or a set communicating observators on this system. The physical variables which can be measured  in this setting are examples of accessible conceptual variables, and are called e-variables in [1].

A maximally accessible variable $\theta^a$ admits values $u_j^a$ that are single eigenvalues of the operator $A^a$, uniquely determined from $\theta^a$. Let $|a;j\rangle$ be the eigenvector associated with this eigenvalue. Then $|a;j\rangle$ can be connected to the question `What is the value of $\theta^a$?' together with the sharp answer `$\theta^a =u_j^a$'. 

A general ket vector $|v\rangle\in\mathcal{H}$ is always an eigenvector of \emph{some} operator associated with a conceptual variable. It is natural to conjecture that this operator quite generally can be selected in such a way that the accessible variable is maximally accessible. Then $|v\rangle$ is in a natural way associated with a question-and-answer pair. It is of  interest that H\"{o}hn and Wever [43] recently derived quantum theory for sets of qubits from such question-and-answer pairs; compare also the present derivation. Note that the interpretation implied by such derivations is relevant for both the preparation phase and  the measurement phase of a physical system.

From a general point of view it may be considered of some value to have an epistemic interpretation which is not necessarily tied to a subjective Bayesian view as it is given in QBism. Under an epistemic interpretation, one may also discuss various ``quantum paradoxes'' like Schr\"{o}dinger's cat, Wigner's friend and the two-slit experiment. 
\bigskip

\textit{Example 1. Schr\"{o}dinger's cat.} The discussion of this example concerns the state of the cat just
before the sealed box is opened. Is it half dead and half alive?

To an observer outside the box the answer is simply: ``I do not know''. Any accessible variable
connected to this observer does not contain any information about the status of life of the cat. But on
the other hand – an imagined observer inside the box, wearing a gas mask, will of course know the
answer. The interpretation of quantum mechanics is epistemic, not ontological, and it is connected to the
observer. Both observers agree on the death status of the cat once the box is opened.
\bigskip

\textit{Example 2. Wigner’s friend.} Was the state of the system only determined when Wigner learned the
result of the experiment, or was it determined at some previous point?

My answer to this is that at each point in time a quantum state is connected to Wigner’s
friend as an observer and another to Wigner, depending on the knowledge that
they have at that time. The superposition given by formal quantum mechanics corresponds to a `do not
know' epistemic state. The states of the two observers agree once Wigner learns the result of the
experiment.
\bigskip

\textit{Example 3. The two-slit experiment.} This is an experiment where all real and imagined observers
communicate at each point of time, so there is always an objective state. 

Look first at the situation when we do not know which slit the particle goes through. This is 
a `do not know' situation. Any statement to the effect that the particles somehow pass through both
slits is meaningless. The interference pattern can be explained by the fact that the particles are (nearly)
in an eigenstate in the component of momentum in the direction perpendicular to the slits in the plane
of the slits. If an observer finds out which slit the particles goes through, the state changes into an
eigenstate for position in that direction. In either case the state is an epistemic state for each of the
communicating observers, real or imagined, and thus also likely to be an ontological state.

 \section{Concluding remarks}
 
 The treatment of this paper is in no way complete. Some open problems include:
 
 - Characterizing exactly when a set of quantum states is in one-to-one correspondence with a set of question-and-answer pairs.
 
 - Giving more concrete conditions under which the Born formula is applicable in practice. This is particularly relevant in connection to cognitive modeling ([28, 29]).
 
 - Developing an axiomatic basis in the spirit of quantum logic (see for instance [68]).
 
 Group theory and quantum mechanics are intimately connected, as discussed in details in [36] for example. In this article it is shown that the familiar Hilbert space formulation can be derived mathematically from a simple basis of groups acting on conceptual variables. The consequences of this is further discussed in [1]. This discussion also seems to provide a link to statistical inference, as indicated in this paper.
 
 From the viewpoint of purely statistical inference the accessible variables $\theta$ discussed in this paper are parameters. In many statistical applications it is useful to have a group of actions $G$ defined on the parameter space; see for instance the discussion in [56]. In the present paper, the quantization of quantum mechanics is derived from the following principle: all model reductions in some given model should be to an orbit (or to a set of orbits) of the group $G$.
 
 It is of some interest that the same criterion can be used to derive the statistical model corresponding to the partial least squares algorithm in chemometrics [37], and also to motivate the more general recently proposed envelope model [38].
 
This paper focuses on group symmetry in spaces of conceptual variables, partucularly those referred to as  parameters/e-variables in [1]. When data are present, in statistical inference one often starts with a group $G^*$ on the data space (see for instance [56]). Referring to the statistical model $P^\theta (X\in B)$ for Borel-sets $B$ in the data-space, the relationship is given by $P^{g\theta}(X\in B)=P^\theta (X\in g^* B)$. This induces a group homomorphism from $G^*$ to $G$. In some cases, this is an isomorphism, and the same symbol $g$ is often used in the data space and the parameter space. However, this is \emph{not} the case when the data space is discrete and the parameter space is continuous, as when the model is given by a binomial distribution or a Poisson distribution. This case is studied in detail in [69]. In these cases, group-theoretical methods were first used to construct the Poisson family and the binomial family, and the basic tool is coherent states for certain groups (the Weyl-Heisenberg group in the Poisson case and the group SU(2) in the binomial case). Finally, inference is studied by reversing the roles of data and parameter. The result in both cases is equivalent to Bayesian inference with a uniform prior on the parameter, which may be a coincidence. Taking this and the present paper as a point of departure, there seems to be a possibility to provide new ideas to symmetry-based approaches to statistical inference. As an extension of [46], a large class of probability distributions are shown to have connections to coherent states in [70]. 
 
 In the present paper, the first axioms of quantum theory are derived from reasonable assumptions. 
 As briefly stated in [1], one can perhaps expect after this that such a relatively simple conceptual basis for quantum theory may facilitate a further discussion regarding its relationship to relativity theory. One can regard physical variables as conceptual variables, inaccessible inside black holes. However, such considerations go far beyond the scopes of both [1] and the present paper. 
 
 Further aspects of the connection between quantum theory and statistical inference theory are under investigation.
 
 \section*{Acknowledgments}
 
 I am grateful to Richard Gill and Ilja Schmelzer for discussions around the Bell theorem, to Gunnar Taraldsen for discussions on fiducial inference, to Anders Rygh Swensen for reference discussions, and to an anonymous  referee on an earlier version of this paper. Also, I want to thank my family for everything they have sacrified while I have been working with theoretical science.
 \bigskip

\section*{References}

\setlength\parindent{0cm}

[1] Helland, I.S.: Epistemic Processes. A Basis for Statistics and Quantum Theory. Springer, Berlin.(2018).

[2] Gill, R.D.: Statistics, causality and Bell's theorem. Statistical Science 20 (4), 512-528 (2014).

[3] Larsson, J.-\AA.: Loopholes in Bell inequality tests of local realism. J. Phys. A: Math. Theor. 47, 424003 (2014).

[4] Giustina, M. et al.: Significant-loophole-free test of local realism. Phys. Rev. Lett. 115, 250401 (2015).

[5] Shalm, L.K. et al.: Strong loophole-free test of local realism. Phys. Rev. Lett. 115, 250402 (2015).

[6] Schmelzer, I.: EPR-Bell realism as a part of logic. arXiv:1712.04334v2 [physics.gen-ph] (2018).

[7] Hossenfelder, S. and Palmer, T.N.: Rethinking superdeterminism. arXiv:1912.06462v2 [quant-ph] (2019).

[8] Helland, I.S.:  Symmetry in a space of conceptul variables.  J. Math. Phys. 60 (5) 052101 (2019). Erratum: J. Math. Phys. 61 (1) 019901 (2020).

[9] Zwirn, H.: Nonlocality versus modified realism. Found. Phys. 50, 1-26 (2020).

[10] Rovelli, C.: Relational quantum mechanics. Int. J. Theor. Phys. 35, 1637 (1996).

[11] Schweder, T. and Hjort, N.L.:  Confidence, Likelihood, Probability. Statistical Inference with Confidence Distributions.  Cambridge University Press. (2016).

[12] Hardy, L.:Quantum theory from reasonable axioms. arXiv: 01010112v4 [quant-ph] (2001).

[13] Chiribella, G., D'Ariano, G.M. and P. Perinotti, P.: Quantum from principles. In: Quantum Theory: Informational Foundation and Foils. Chiribella, G. and Spekkens, P.W.  [Eds.] pp. 171-221. Springer, Berlin (2016).

[14] Daki\'{c}, B. and Brukner, \v{C}.: Quantum theory and beyond: Is entaglement spacial? arXiv:09110695v1 [quant-ph] (2009).

[15] Goyal, P.: Information-geometric reconstruction of quantum theory. Phys. Rev. A 78, 052120 (2008).

[16] Masanes, L. and M\"{u}ller.: A derivation of quantum theory from physical requiremrnts. New J. Phys, 13, 063001 (2011).

[17] G. Chiribella, G., Cabello,A., Kleinmann, M. and M\"{u}ller, M.P.: General Bayesian theories and the emergence of the exclusitivity principle. arXiv: 1901.11412v2 [quant-ph] (2019).

[18] Daki\'{c}, B. and Brukner, \v{C}.: The classical limit of a physical theory and the dimensionality of space. arXiv:1307.3984v1 [quant-ph] (2013).

[19] Robinson, M.: Symmetry and the Standard Model. Springer, New York (2011).

[20] Susskind, L. and Friedman, A.: Quantum Mechanics. The Theoretical Minimum. Basic Books, New York. (2014).

[21] Schlosshauer, M., Kofler, J. and Zeilinger, A.: A snapshot of fundamental attitudes towards quantum mechanics. Sudies in History and Philosophy of Modern Physics 44, 222-238. (2013).

[22] Norsen, T. and Nelson, S.: Yet another snapshot of fundamental attitudes toward quantum mechanic. arXiv: 1306.4646v2 [quant-ph]. (2013).

[23] Fuchs, C.A.:  QBism, the perimeter of quantum Bayesianism arXiv: 1003.5209 [quant-ph]. (2010).

[24] Fuchs, C.A.: Quantum mechanics as quantum information (and only a little more). In: Ed. Khrennikov, A.: Quantum Theory: Reconstruction of Foundation. V\"{a}xj\"{o} Univ. Press, V\"{o}xj\"{o}. quant-ph/0205039 (2002).

[25] Fuchs, C.A. and Schack, R.: Quantum-Bayesian Coherence. arXiv:0906.2187 [quant-ph] (2009).

[26] von Baeyer, H.C.: QBism: The future of quantum physics. Harvard University Press, Harvard.(2016).

[27] Zwirn, H.: Is QBism a possible solution to the conceptual problems of quantum mechanics? Preprint (2019).

[28] Pothos, E.M. and  Busemeyer, J.R.: Can quantum probability provide a new direction for cognitive modeling? With discussion. Behaviorial and Brain Sciences 36, 255-327. (2013).

[29] Busemeyer, J.R. and Buza, P.D.: Quantum Models for Cognition and Decision. Cambridge University Press, Cambridge. (2012).

[30] Helland, I.S.: Steps Towards a Unified Basis for Scientific Models and Methods. World Scientific, Singapore.(2010).

[31] Nachbin, L.: The Haar Integral.  Van Nostrand, Princeton, NJ. (1965).

[32] Hewitt, E. and Ross, K.A.:  Abstract Harmonic Analysis, II. Springer-Verlag, Berlin. (1970).

[33] Wijsman, R.A.: Invariant Measures on Groups and Their Use in Statistics.  Lecture Notes - Monograph Series 14, Institute of Mathematical Statistics, Hayward, California. (1990).

[34] Perelomov, A.:  Generalized Coherent States and Their Applications.  Springer-Verlag, Berlin. (1986).

[35] Berger, J.O.: Statistical Decision Theory and Bayesian Inference, 2nd Ed.. Springer-Verlag, New York (1985).

[36] Hall, B.C.:  Quantum Theory for Mathematicians.  Graduate Texts in Mathematics, 267, Springer, Berlin. (2013).

[37] Helland, I.S., S\ae b\o, S. and Tjelmeland, H.:  Near optimal prediction from relevant components.  Scandinavian Journal of Statistics 39, 695-713. (2012).

[38] Helland, I.S., S\ae b\o, S., Alm\o y, T. and Rimal, R.:  Model and estimators for partial least squares.  Journal of Chemometrics 32:  e3044. (2018).

[39] Gazeau, J.-P.: Coherent States in Quantum Physics. Wiley-VCH, Weinberg (2009).

[40] Ma, Z.-Q.: Group Theory for Physicists. World Scientific, Singapore (2007).

[41] Helland, I.S.:  When is a set of questions to nature together with sharp answers to those questions in one-to-one correspondence with a set of quantum states? arXiv: 1909.08834 [quant-ph] (2019).

[42] Plotnitsky, A.: Niels Bohr and Complementarity. An Introduction. Springer, New York. (2013).

[43] Khrennikov, A.: Ubiquitous Quantum Structure. From Psychology to Finance. Springer, Berlin. (2010).

[44] Yukalov, V.I. and Sornette, D.: Quantum decision theory as a quantum theory of measurement. Phys. Lett. A  372, 6867-6871 (2008).

[45] Yukalov, V.I. and Sornette, D.: Processing information in quantum decision theory. Entropy 11, 1073-1120 (2009).

[46] Yukalov, V.I. and Sornette, D.: Mathematical structure of quantum decision theory. Adv. Compl. Syst. 13, 659-698 (2010).

[47] Yukalov, V.I. and Sornette, D.: Decision theory with prospect interference and entanglement. Theory Dec. 70, 383-328 (2011).

[48] Yukalov, V.I. and Sornette, D.: How brains make decisions. Springer Proceedings in Physics 150, 37-53 (2014).

[49] Caves, C.M., Fuchs, C.A., Manne, K. and Renes, J.M.: Gleason-type derivations of the quantum probability rule for generalized measurements. arXiv:036179v1 [quant-ph] (2003).

[50] Sharpnel, S., Costa, F. and Milburn, G.: Updating the Born rule. arXiv:1702.01845v1 [quant-ph] (2017).

[51] Busch, P.: Quantum states and generalized observables. A simple proof of Gleason's Theorem. Phys. Rev. Lett. 91 (12), 120403.(2003).

[52] Lehmann, E.L. and Casella, G.: Theory of point estimation. Springer, New York (1998).

[53] Taraldsen, G. and Lindqvist, B.H.: Improper priors are not improper.  The American Statistician 64 (2), 154-158.

[54] Bickel, P.J. and Doksum, K,A.: Mathematical Statistics. Basic Ideas and Selected Topics. Prentice Hall, New Jersey. (2001).

[55] Ferguson, T.S.: Mathematical Statistics: A Decision Theoretic Approach. Academic Press, Cambridge, MA. (1967).

[56] Helland, I.S.:  Statistical inference under symmetry.  International Statistical Review 72, 409-422. (2004).

[57] Taraldsen, G. and Lindqvist, B.H.:  Fiducial theory and optimal inference. Annals of Statistics 41 (1), 323-341.

[58] Fisher, R.A.: Inverse probability. Math. Proc. Cambridge Philos. Soc. 26, 528-535. (1930).

[59] Fraser, D.A.S.: On fiducial inference. Annals of Mathematical Statistics 32 (3), 661-676.

[60] Fraser, D.A.S.: The fiducial method and invariance. Biometrika 48 (3/4), 261-280.

[61] Taraldsen, G. and Lindqvist B.H.: Fiducial and posterior sampling. Communications in Statistics - Theory and Methods 44, 3754-3767. (2015).

[62] Vedral, V.: Living in a quantum world. Scientific American 304 (6), 20-25. (2011).

[63] D. Aerts, D., Sozzo, S. and Tapia, J.. Identifying quantum structures in the Ellsberg paradox. Intern. J. Theor. Phys. 53, 3666-3682. (2014).

[64] Jacobs, K.: Quantum Measurement Theory and Its Applications. Cambridge University Press, Cambridge. (2014).

[65] Wigner, E.P.: Group theory and its application to the quantum mechanics of atomic spectra. Academic, New York. (1959).

[66] Fuchs, C.A.: QBism, the perimeter of quantum Bayesianism. arXiv: 1003.5209 [quant-ph] (2010).

[67] H\"{o}hn, P.A. and Wever, C.S.P.: Quantum theory from questions. Phys. Rev. A 95, 012102 (2017).

[68] Hughes, R.I.G.: The Structure and Interpretation of Quantum Mechanics. Harvard University Press, Cambridge, Mass. (1989).

[69] Heller, B. and Wang, M.:  Group invariant inferred distributions via noncommutative probability.  Recent Developments in Nonparametric Inference and Probability, IMS Lecture Notes - Monograph Series 50, 1-19. (2006).

[70] Ali, S.T. Gazeau, J.-P. and Heller, B.:  Coherent states and Bayesian duality.  J. Phys. A: Math. Theor. 41, 365302. (2008).

\end{document}